\documentclass[12pt, notitlepage,aps,preprintnumbers,superscriptaddress,nofootinbib,aps,prd,floatfix]{revtex4-2}
\usepackage{physics}
\usepackage{bm}
\usepackage{graphicx}
\usepackage{epstopdf}
\usepackage{hyperref}
\usepackage{multirow}
\usepackage{array}
\usepackage{xcolor}
\usepackage[draft]{todonotes}
\usepackage{wrapfig}
\usepackage{ifthen}
\usepackage{pdfcomment}
\usepackage{adjustbox}
\usepackage{listings}
\usepackage{version}
\usepackage{amssymb}
\usepackage{booktabs}
\usepackage{makecell}
\usepackage{enumitem}
\setlist{nolistsep}
\definecolor{nicered}{rgb}{0.5,0.,0.}
\definecolor{nicegreen}{rgb}{0.,0.5,0.}
\definecolor{niceblue}{rgb}{0.,0.,0.5}
\hypersetup{colorlinks,citecolor=nicegreen,linkcolor=nicered,urlcolor=niceblue}
\usepackage{orcidlink, xspace}

\usepackage{ulem}

\begin{document}

\title{Impact of lattice gluon dataset on CTEQ-TEA global PDFs}

\author{Alim Ablat\,\orcidlink{0009-0005-1987-7755}}
\email{alimjanablat@outlook.com}
\affiliation{School of Physics and Electrical Engineering, Kashi University, Kashi, Xinjiang 844000 China\looseness=-1}

\author{William Good\, \orcidlink{0000-0001-8428-1003}}
\email{goodwil9@msu.edu}
\affiliation{Department of Physics and Astronomy, Michigan State University, East Lansing, Michigan 48824, USA\looseness=-1}
\affiliation{Department of Computational Mathematics,
  Science and Engineering, Michigan State University, East Lansing, Michigan 48824, USA}
  
\author{Sayipjamal Dulat\,\orcidlink{0000-0003-2087-0727}}
\email{sdulat@hotmail.com}
\affiliation{School of Physics Science and Technology, Xinjiang University, Urumqi, Xinjiang 830046 China\looseness=-1}
\affiliation{Department of Physics and Astronomy, Michigan State University, East Lansing, Michigan 48824, USA\looseness=-1}

\author{Tie-Jiun Hou\,\orcidlink{0000-0002-5064-1012}}
\email{tjhou@msu.edu}
\affiliation{School of Nuclear Science and Technology, University of South China, Hengyang, Hunan 421001, China}

\author{Huey-Wen Lin\,\orcidlink{0000-0001-6281-944X}}
\email{hueywen@msu.edu}
\affiliation{Department of Physics and Astronomy, Michigan State University, East Lansing, Michigan 48824, USA\looseness=-1}

\author{Keping Xie\,\orcidlink{0000-0003-4261-3393}}
\email{xiekepi1@msu.edu}
\affiliation{Department of Physics and Astronomy, Michigan State University, East Lansing, Michigan 48824, USA\looseness=-1}

\author{C.-P. Yuan\,\orcidlink{0000-0003-3988-5048}}
\email{yuanch@msu.edu}
\affiliation{Department of Physics and Astronomy, Michigan State University, East Lansing, Michigan 48824, USA\looseness=-1}

\preprint{MSUHEP-25-001}

\begin{abstract}
We investigate the potential impact of the gluon parton results from lattice QCD on the global parton distribution function  analysis within the CTEQ-TEA framework.
The dependence on PDF parameterization is explored using the CT18As variant, incorporating the ATLAS 7-TeV precision $W, Z$ dataset and introducing more flexible parameters to allow strangeness asymmetry at the starting scale $Q_0$.
The interplay between lattice input and jet datasets is examined by including the post-CT18 inclusive jet datasets from recent Large Hadron Collider (LHC) measurements and/or removing all-inclusive jet datasets.
Finally, we demonstrate several potential phenomenological implications for the LHC, focusing on gluon-gluon parton luminosity and related processes, such as the production of a Higgs-like scalar, top-quark pairs, and their associated production with an additional jet, Higgs, or Z~boson.
\end{abstract}

\maketitle

\section{Introduction}
\label{sec:intro}

As the mediator of the strong interaction, the gluon plays a crucial role in the internal structure of hadrons at both the perturbative and nonperturbative levels of Quantum Chromodynamics (QCD).
Unlike quarks, gluons carry adjoint color charge and directly self-interact, leading to complex QCD dynamics.
At the perturbative level, gluons drive the evolution of parton distribution functions (PDFs) through the Dokshitzer-Gribov-Lipatov-Altarelli-Parisi (DGLAP) equations~\cite{Gribov:1972ri,Lipatov:1974qm,Altarelli:1977zs,Dokshitzer:1977sg}.
In modern hadron colliders, the gluon PDF plays a central role in high-energy hadronic collisions, where gluon-initiated processes often dominate.
Understanding the gluon PDF is essential for interpreting key LHC measurements, such as jet production, Higgs-boson production, and top-quark pair production.
These processes are critical for both Standard Model (SM) precision tests and searches for physics beyond the Standard Model (BSM), including high-mass resonance studies.

Since the gluon is not directly observable in deep inelastic scattering (DIS), its distribution is inferred from scaling violations~\cite{Bjorken:1968dy,Gribov:1972ri,Lipatov:1974qm,Altarelli:1977zs,Dokshitzer:1977sg} and measurements of hadronic final states.
Consequently, the precise determination of the gluon PDF remains a cornerstone of modern QCD research.
Recent global fits~\cite{Hou:2019efy,Bailey:2020ooq,NNPDF:2021njg} incorporated next-to-next-to-leading order (NNLO) QCD corrections and high-precision experimental datasets from DIS, Drell-Yan pair production, and jet and heavy-quark production at the Tevatron and the LHC.
These advancements have significantly enhanced our understanding of the gluon PDF, providing a robust theoretical foundation for both current and future collider experiments.
However, challenges persist, particularly in the small- and large-$x$ regions, where experimental measurements are either sparse or absent.
Advancing our understanding of the gluon PDF in these extreme-$x$ regions remains a key objective.
In this work, we focus specifically on the large-$x$ gluon distribution.

Lattice QCD is a non-perturbative theoretical method for calculating QCD quantities that can have full systematic control.
The development of the Large-Momentum Effective Theory (LaMET)~\cite{Ji:2013dva,Ji:2014gla,Ji:2017rah} has enabled calculations of $x$-dependent hadron structure, particularly in the large-$x$ region, using nonperturbative lattice-QCD simulations.
Numerous lattice studies have investigated nucleon and meson PDFs, along with generalized parton distributions (GPDs), using the LaMET approach (also known as the ``quasi-PDF'' method) and the pseudo-PDF framework~\cite{Radyushkin:2017cyf,Orginos:2017kos,Zhang:2018diq,Li:2018tpe}.
Furthermore, alternative methods such as the Compton amplitude approach~\cite{Martinelli:1998hz,Horsley:2012pz,Chambers:2014qaa,Chambers:2015bka,Chambers:2017dov}, the ``hadronic tensor'' method~\cite{Liu:1993cv,Liu:1999ak}, and the ``Lattice Cross Sections'' approach~\cite{Ma:2017pxb} have been developed.
For a detailed discussion of these techniques, we refer the reader to recent review articles~\cite{Gao:2024pia,Lin:2023kxn,Amoroso:2022eow,Ji:2020ect,Ethier:2020way,Constantinou:2020hdm,Lin:2017snn}.

While considerable progress has been made in calculating isovector quark PDFs, the gluon PDF remains less explored due to the notoriously noisy matrix elements in lattice calculations.
To date, only a few exploratory studies of the unpolarized gluon PDF have been conducted for nucleon~\cite{Fan:2018dxu,Fan:2020cpa,HadStruc:2021wmh,Fan:2022kcb,Good:2023gai,Delmar:2023agv,Good:2024iur,Chowdhury:2024ymm}, pion~\cite{Fan:2021bcr,Good:2023ecp}, and kaon~\cite{Salas-Chavira:2021wui}, as well as for the polarized nucleon~\cite{HadStruc:2022yaw}, using the pseudo-PDF~\cite{Balitsky:2019krf} and quasi-PDF~\cite{Zhang:2018diq,Wang:2019tgg} methods.
Many exploratory lattice studies rely on a single lattice spacing and a heavy pion mass.
However, the MSULat group's pioneering calculation of the nucleon gluon PDF incorporates a continuum-physical extrapolation~\cite{Fan:2022kcb,Good:2023gai}, marking a significant advancement in controlling lattice artifacts in gluon-PDF calculations.

MSULat group reported the only continuum-limit study of the nucleon unpolarized gluon PDF\footnote{During the referee process of this manuscript, two additional continuum-limit studies using quasi-PDF methods have been reported by the MSULat and LPC groups~\cite{NieMiera:2025mwj,NieMiera:2025vcx,ChenChen:2025amm}.} using three lattice spacings, 0.09, 0.12, and 0.15~fm with three pion masses $M_\pi \approx 220$, 310, and 690~MeV, using the pseudo-PDF method to obtain lightcone PDFs from Ioffe-time distributions (ITDs)~\cite{Fan:2022kcb}.
High-statistics measurements and Gaussian momentum smearing~\cite{Bali:2016lva} are used on the quark fields to improve the signal for nucleon boost momenta up to 2.56~GeV.
Multiple source-sink separations are used to extract ground-state matrix elements from the ``two-sim'' fit to ensure that excited-state contamination is removed~\cite{Fan:2022qve}.
In Ref.~\cite{Good:2023rbp}, the effects of smearing on the elements of the gluon matrix were investigated using different smearing methods at various steps.
The study found that the MSULat group's smearing choice exhibited a milder effect than Stout smearing or Wilson flow, as employed by other groups~\cite{HadStruc:2021wmh,Delmar:2023agv}.
The dependence of the lattice spacing of the matrix elements with the pion mass at 310 and 700~MeV is studied, checking both the $O(a)$ and $O(a^2)$ forms using multiple continuum extrapolation strategies.
In the end, a physical continuum extrapolation is performed to obtain continuum-reduced pseudo-ITDs (RpITDs), before the nucleon unpolarized gluon PDF $xg(x)$ is finally obtained from the $xg(x)/\langle x \rangle_g$ and $\langle x \rangle_g$ results.
The final gluon PDF is obtained with the gluon momentum fraction input calculated on the same ensemble~\cite{Fan:2022qve}.

Although the current lattice gluon PDF calculation does not yet meet the latest Flavour Lattice Averaging Group (FLAG) standards~\cite{FlavourLatticeAveragingGroupFLAG:2024oxs}, it is natural to explore the synergy and complementarity between lattice-QCD results and global PDF fits, particularly given the scarcity of experimental measurements in the large-$x$ region.
The interplay between lattice input and jet datasets is examined by including the post-CT18 inclusive jet datasets from recent measurements at the Large Hadron Collider (LHC) and/or removing all-inclusive jet datasets.
Finally, we discuss several potential phenomenological implications at the LHC, focusing on gluon-gluon parton luminosity and related processes, such as the production of a Higgs-like scalar, top-quark pairs, and their associated production with an additional jet, Higgs, or Z boson.
We must estimate the precision needed in the lattice calculation to assess what computational resources are needed to make desired phenomenological impacts that would be interesting to the wider physics community.
A similar synergy between lattice and global fits has been explored in the past in the non-singlet quark PDFs.
For example, the JAM Collaboration conducted a study on pion PDFs utilizing PDF inputs of lattice valence-quark~\cite{JeffersonLabAngularMomentumJAM:2022aix}.
However, there has yet to be a study of the potential impact on global PDF fits using lattice-QCD gluon-related calculations.\footnote{During the review process for this manuscript, a new study by the MSULat and JAM collaborations investigating the impact of lattice data on the pion gluon PDF was reported in Ref.~\cite{Good:2025nny}.}

Building on the approach of our previous work on lattice strangeness asymmetry~\cite{Hou:2022onq}, we extend this methodology to incorporate the latest lattice gluon PDF~\cite{Fan:2022kcb,Good:2023gai} into the CT18 global-analysis framework~\cite{Hou:2019efy,Hou:2022onq,Ablat:2024uvg}.
This integration leverages the strengths of both lattice simulations and global PDF fits, enhancing our understanding of the gluon PDF in the large-$x$ region.
This paper is structured as follows.
Section~\ref{sec:gQCDa} provides the methodology of the CTEQ-TEA global QCD analysis framework of PDFs that we use as a  baseline fit for the comparisons, Bjorken-$x$ lattice data to be included in this work and how we incorporate lattice-QCD calculations of the gluon PDF data within the CT18 framework~\cite{Hou:2019efy,Hou:2022onq,Ablat:2024uvg}.
In Sec.~\ref{Sec:CT18+LattGluon}, we study the impact of lattice-QCD data  on the central PDFs, associated error bands, and its interplay with collider inclusive jet datasets.
We also discuss potential phenomenological implications for the future improved gluon PDFs in Sec.~\ref{sec:pheno}. 
Finally, we explore alternative lattice-QCD inputs by incorporating pseudo-PDF matrix elements alongside experimental cross-sections in Sec.~\ref{Sec:Analysis2} and 
conclude in Sec.~\ref{Sec:Conclusion}.

\section{Methodology}\label{sec:gQCDa}

A global analysis of PDFs of the proton is a systematic approach to extract the probability distributions of quarks and gluons (collectively known as partons) within the proton as a function of their momentum fraction $x$ and the energy scale $Q^2$.
This process involves several steps, which can be outlined as follows:

{\bf Data Collection.}
The first step involves gathering a comprehensive set of experimental data from various high-energy physics experiments. This data typically includes measurements from DIS, proton-proton collisions, and other processes that are sensitive to the structure of the proton. Key experimental collaborations include HERA, the LHC, and fixed-target experiments.

PDFs of the proton are essential inputs for calculating high-energy cross sections involving initial-state hadrons.
The universal PDFs are extracted from the global analysis of parton distributions by incorporating experimental data from a wide range of physical processes and by the use of the Dokshitzer–Gribov–Lipatov–Altarelli–Parisi (DGLAP) evolution equations.
In this section, we discuss the essential steps and how the lattice-QCD data is incorporated.

In this work, the baseline analysis considers the same selected experimental data sets as those in the CT18 global analysis~\cite{Hou:2019efy}.
Below, we briefly list the included physical processes and corresponding experiments,
given by their commonly used abbreviations in the CTEQ-TEA PDF global analysis:
\begin{itemize}
\item Deep inelastic scattering (DIS): HERA I+II, BCDMS, NMC, CDHSW, CCFR, NuTeV, H1, HERA charm;
\item Drell–Yan production: E605, E866, CDF, D0, ATLAS, CMS, LHCb;
\item Top-quark pair production: ATLAS, CMS;
\item Inclusive jet production: CDF, D0, ATLAS, CMS.
\end{itemize}
For more detailed information, we refer readers to Tables~I and II of the original CT18 global-analysis work~\cite{Hou:2019efy}.

\begin{figure}[tb!]
\centering
\includegraphics[width=0.49\linewidth]{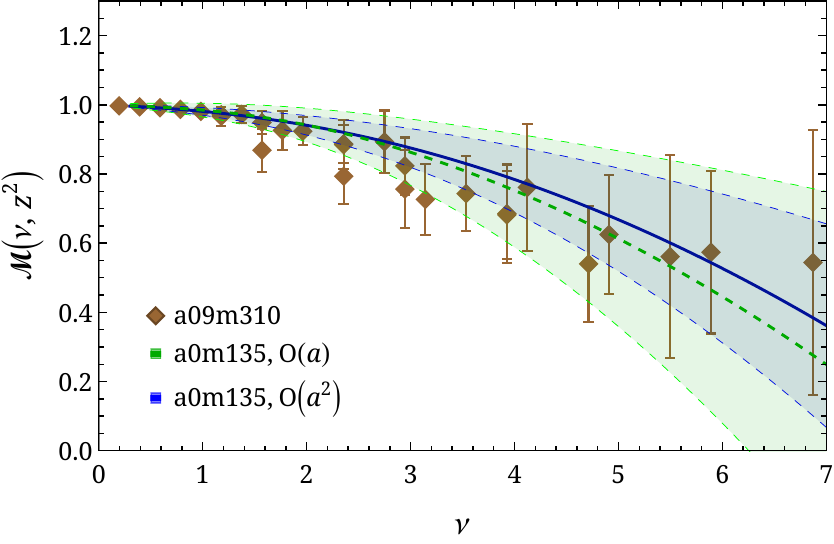}
\includegraphics[width=0.49\linewidth]{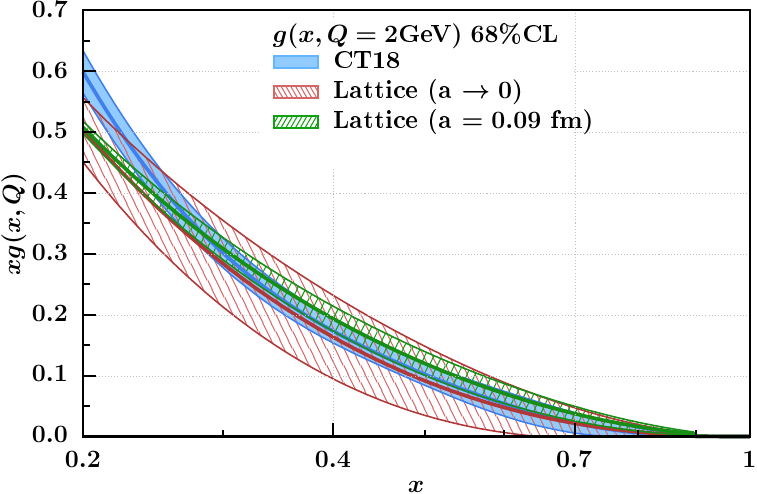}
\caption{
The RpITDs from 310-MeV with lattice spacing $a\approx 0.09$~fm lattice ensemble (brown points), and
the physical-continuum limit extrapolated from all ensembles in Ref.~\cite{Fan:2022kcb}
with $a$ (green band) and $a^2$ (blue band) continuum extrapolation at the physical pion mass.
Right: The unpolarized gluon PDF, $xg(x,Q)$, a function of $x$ with $x \in [0,1]$ obtained from CT18 (blue), MSULat's continuum-physical (red), and a09m310-ensemble (green) results; note that they are consistent within statistical errors.
The error bands are displayed at the 68\% confidence level (CL).}
\label{fig:LatGluon}
\end{figure}

\begin{figure}[!h]
    \centering
    \includegraphics[width=0.49\linewidth]{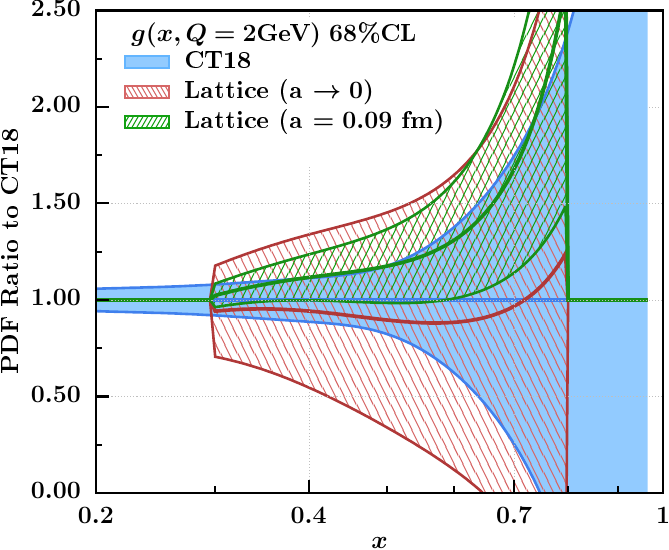}
    \includegraphics[width=0.49\linewidth]{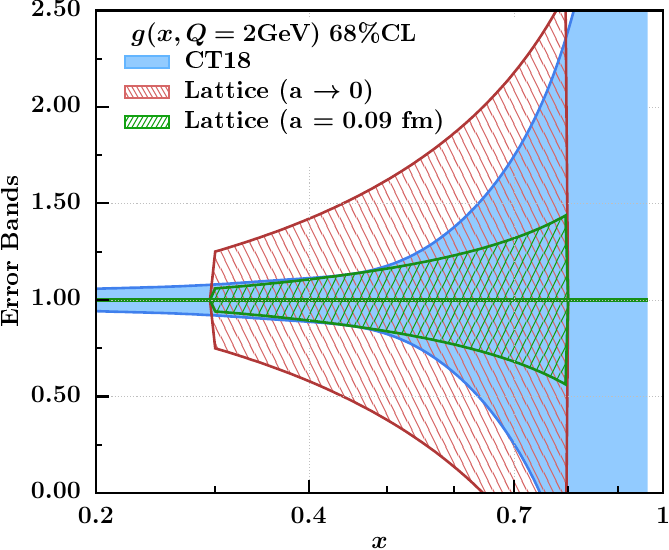}
    \caption{Comparison of the central values and errors of the gluon PDF from the CT18 fit, lattice gluon PDF at the physical continuum limit ($a\rightarrow 0$), and a lattice spacing of
    $a = 0.9$~fm for $Q = 2$~GeV.}
    \label{fig:GluonPDFs}
\end{figure}

{\bf Data from Lattice-QCD Calculation.}
This study aims to take the lattice calculation of $x$-dependent gluon PDF from pseudo-PDF method as additional
data to explore its impact on the baseline, i.e., CT18, analysis.   
The results of two lattice gluon PDF calculations will be separately considered in addition to the above-mentioned experimental data sets.
One is to consider the lattice gluon PDF calculated at the physical continuum limit ($a \rightarrow 0$ with pion mass $M_\pi \approx 135$~MeV), and another is to consider a single ensemble $M_\pi\approx 310$~MeV with a lattice spacing $a \approx 0.09$~fm from the MSULat group~\cite{Fan:2022kcb}.
In Fig.~\ref{fig:LatGluon}, we observe that both results are consistent within statistical uncertainties for the lattice double-ratio matrix elements (left panel) and the extracted gluon PDF (right panel).
For the remainder of this paper, we will use the single-ensemble result at the finer lattice spacing of $a \approx 0.09$~fm as a proxy for the targeted continuum-physical precision achievable by lattice QCD calculations soon. This will serve as the basis for studying both current and anticipated future lattice impacts.
The left-hand side of Fig.~\ref{fig:GluonPDFs} compares CT18 to both lattice gluon PDF determinations by normalizing everything by the best CT18 gluon fit, while the right-hand side emphasizes the relative errors of each method by normalizing each fit by its best-fit value.
We compare the fitted gluon PDF to the lattice gluon PDF at $Q = 2$~GeV, which is the energy scale at which lattice results are typically presented.
We select the lattice gluon PDF inputs conservatively in the regions $x \in [0.3, 0.8]$ or $[0.4, 0.7]$.
These regions are chosen because the lattice-determined PDFs exhibit smaller systematic uncertainties related to the boost momentum, approximately 2.6~GeV, and existing experimental datasets do not strongly constrain the current global fits.
Each case is treated independently as a new dataset within the comprehensive collection of data used in the CT18 fits.

{\bf Theoretical Framework.}
The theoretical predictions for the experiments included in the PDF global analysis are based on QCD and electroweak theory.
In the analysis, the cross-sections for various scattering processes can be calculated using perturbative QCD, which describes the interactions of quarks and gluons at high energies.
The theory calculations also depend on the PDFs, which encapsulate the nonperturbative aspects of the proton's structure.

Based on the QCD factorization theorem, a typical cross section for hadron-hadron scattering is given by summing over different partonic sub-processes that contribute to the final state under consideration, convoluted with the respective PDFs at the relevant hard scale as
\begin{equation}
\sigma_{h_1,h_2} (p_{h_1}, p_{h_2}) = \sum_{i,j}\int_{0}^{1}\dd x_i \int_{0}^{1}\dd x_j f_{i/{h_1}}(x_i,\alpha_s,\mu^2_F)
f_{j/{h_2}}(x_j,\alpha_s,\mu^2_F) \hat \sigma_{{h_1},{h_2}} (p_i,p_j,\mu^2_R),
\end{equation}
where $i, j \in {g, u, d, c, s, b}$; the $p_{h_1}$ and $p_{h_2}$ are the parent hadrons four-momenta;
$p_i = x_i p_{h_1}(p_j = x_j p_{h_2})$ is the four-momentum of the incoming parton $i(j)$ in the hard scattering;
$\mu_R$ and $\mu_F$ are the renormalization scale and factorization scale that arise in $\alpha_s(\mu_R)$ and the PDFs, respectively;
$f_{i/h_1(j/h_2)}(x_i,\alpha_s,\mu^2_F)$ is the distribution of a parton $i(j)$ with momentum fraction $x_{i(j)}$ at the scale $\mu_F$;
$\hat \sigma_{{h_1},{h_2}} (p_i,p_j,\mu^2_R,\mu^2_F)$ is the hard cross section, computed order-by-order in $\alpha_s(\mu_R)$ at the scale $\mu_R$;  

To ensure consistency, the hard cross sections and the evolution kernels must be computed at the same order in perturbative QCD.
In the current CTEQ-TEA global QCD analysis, the  QCD calculation is carried out up to the next-to-next-to-leading order (NNLO).

{\bf Parametrization of Nonperturbative PDFs.}
To perform a global QCD analysis, one must specify a set of initial parton distribution functions at the starting scale $\mu = Q_0$, and evolve them to a higher energy scale $Q$ using the Dokshitzer–Gribov–Lipatov–Altarelli–Parisi (DGLAP) evolution equations~\cite{Gribov:1972ri,Lipatov:1974qm,Altarelli:1977zs,Dokshitzer:1977sg}.
These equations describe how the PDFs change with the energy scale due to the splitting of partons into other partons.
In the CT18 fits, the initial scale $Q_0$ is chosen to be 1.3~GeV.
The general functional form of the nonperturbative PDFs at the initial scale $Q_0$, expressed in terms of free parameters $a_k$, is given by
\begin{equation}
f_i(x,Q=Q_0) = a_0 x^{a_1-1} (1-x)^{a_2} P_i(y,a_3,a_4,\ldots),
\end{equation}
where $y=\sqrt{x}$, and the coefficients $a_1$ and $a_2$ control the asymptotic behavior of $f_i(x, Q_0)$ in the limits $x \to 0$ and $x \to 1$, respectively.
For the extraction of the CT18 series PDFs~\cite{Hou:2019efy,Hou:2022onq}, we use Bernstein polynomials (also called a ``B\'ezier curve"~\cite{Kotz:2023pbu}) for the function $P_i(y, a_3, a_4, \ldots)$.
More details can be found in Appendix~C of Ref.~\cite{Hou:2019efy}.

The gluon PDF exhibits significant uncertainties at both small and large values of $x$.
In this paper, we focus on the gluon distribution.
As an example, we present below the parameterization used for the gluon PDF in the CT18 fits:
\begin{equation}\label{eq:ParamGluon}
g(x,Q=Q_0) = a_0 x^{a_1-1}(1-x)^{a_2}\,P^g_a(y),
\end{equation}
with
\begin{equation}
P^g_a(y) = \sinh(a_3) (1-y)^3 + \sinh(a_4) 3 y (1-y)^2 + a_5 3 y^2 (1-y) + y^3, \nonumber
\end{equation}
where $a_5$ is related to $a_1$ as $a_5 = (3 + 2a_1)/3$.
The $Q$ dependence of $f(x,Q)$ is obtained by solving the QCD DGLAP evolution equations.

{\bf Statistical Analysis and Fitting Procedure.}
A statistical analysis method, often based on a chi-squared minimization approach, is employed to compare the theoretical predictions with the experimental data. The fitting procedure involves:

\begin{itemize}

\item {\it Defining a chi-squared function:} This function quantifies the difference between the observed data and the theoretical predictions, weighted by the uncertainties in the measurements.
\item  {\it Minimization:}
The parameters of the chosen PDF functional forms are varied to minimize the chi-squared value.
\item {\it Uncertainty Estimation:}
After obtaining the best-fit parameters, the uncertainties in the PDFs are estimated using techniques like the Hessian method~\cite{Pumplin:2001ct,Pumplin:2002vw} or Monte Carlo replicas~\cite{Forte:2010dt,Forte:2013wc}.
\end{itemize}

In the CTEQ-TEA global analysis,
the best-fit parameters of the PDFs are determined by calculating the $\chi^2$ between the theoretical predictions $T_k(a_i)$, which depend on a set of PDF parameters $a_i$ ($i\in [1, n]$), and the experimental data $D_k$ ($k\in[1,N]$), after taking into account the statistical uncertainties, uncorrelated systematic errors, and correlated systematic errors.
Specifically, the following global $\chi^2$ function is used to determine the best-fit parameters by minimizing $\chi^2$ through adjustments to the parameterizations of the input PDFs.

\begin{equation}
\chi_{E}^{2}(a,\lambda)=\sum_{k=1}^{N_\text{pt}}\frac{1}{s_{k}^{2}}\left(D_{k}-T_{k}(a)-\sum_{\alpha=1}^{N_{\lambda}}\lambda_{\alpha}\beta_{k\alpha}\right)^{2}+\sum_{\alpha=1}^{N_{\lambda}}\lambda_{\alpha}^{2},\label{Chi2sys}
\end{equation}
where
$s_{k}\equiv \sqrt{s_{k,\text{stat}}^{2}+s_{k,\text{uncor sys}}^{2}}$
is the total uncorrelated error associated with the measurement $D_k$, and $\beta_{k\alpha}$ represents the correlated systematic errors.
It is common practice to associate each source of correlated uncertainty with an independent nuisance parameter $\lambda_{\alpha}$, which is assumed to be sampled from a standard normal distribution, unless known otherwise.

Various methods exist to quantify uncertainties in PDFs, such as the Hessian method~\cite{Pumplin:2001ct, Pumplin:2002vw}, the Monte-Carlo approach~\cite{Giele:1998gw,Giele:2001mr}, and the Lagrange multiplier technique~\cite{Stump:2001gu}.
Our default PDF set in this study is CT18, which employs the Hessian method to estimate the corresponding uncertainties~\cite{Hou:2019efy}.
The PDF induced uncertainty of any variable (including experimental observables) that depends on the PDFs, can be estimated using the master formulas provided in Refs.~\cite{Nadolsky:2001yg,Pumplin:2002vw}:
\begin{equation}\label{master}
\Delta X = \frac{1}{2} \left(\sum_{i=1}^{N_\text{ev}} \left[ X\left(f_{i}^{+}\right) - X\left(f_{i}^{-}\right) \right]^{2} \right)^{1/2}
\end{equation}
where $X\left(f_{i}^{\pm}\right)$ can be a physical quantity, or a particular PDF such as the gluon distribution at some specific $x$ and $Q$ values, or even one of the PDF parameters $a_i$;
and $N_\text{ev}$ is the number of eigenvectors.
More details can be found in Sec.~III of Ref.~\cite{Hou:2019efy}.

\section{Impact of lattice gluon input on CTEQ-TEA PDFs} \label{Sec:CT18+LattGluon}

In this section, we analyze how the proton PDFs are modified by incorporating lattice-QCD calculations of the gluon PDF as an additional dataset in the CT18 analysis. We also assess its impact relative to other data sources, such as DIS, Drell-Yan, jets, and top-quark pairs, included in the original CT18 fit.
The global fitting procedure follows the standard configuration of the CT18 analysis, as discussed in the previous section. We maintain the same tolerance criterion for estimating PDF uncertainties while adopting the non-perturbative parametrization forms used in the CT18 fits at the initial scale,  $Q_0=1.3$~GeV.
The inclusion of lattice-QCD calculations, which constrain the gluon PDF at large $x$, necessitates a more detailed examination of the PDF parametrization. Developing more flexible functional forms for the gluon PDF requires significant effort and will be explored in future research.

\subsection{Constraints on CT18 and CT18As PDFs at large-$x$}
\label{sec:impact}

We consider two lattice inputs of the gluon PDF at the scale $Q=2$~GeV, which are provided with distinct lattice spacings, $a \to 0$ (physical continuum limit) and $a \approx 0.09$~fm.
In both scenarios, we explore a conservative choice of momentum-fraction range $x \in [0.3, 0.8]$ as well as an alternative one $x \in [0.4, 0.7]$ to estimate the systematics, and find the physical-continuum limit dataset $(a\to 0$) play a negligible role in the global analysis.
It can be understood in terms of the large uncertainty from the continuum-limit extrapolation after including the lattice discretization systematics, as shown in Fig.~\ref{fig:LatGluon}.
For this reason, we will mainly focus on the lattice input with the spacing $a \approx 0.09$~fm, whose PDF is consistent within the current statistical errors, in the rest of this work, as a future targeted precision for the lattice community.

\begin{figure}[!h]
    \centering
    \includegraphics[width=0.49\linewidth]{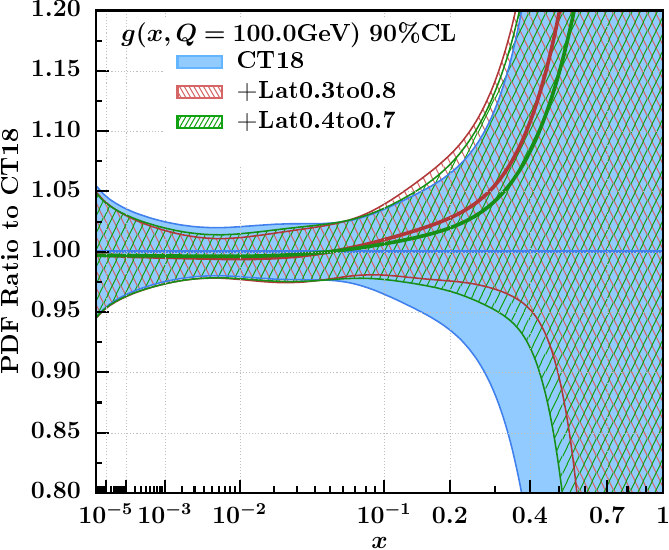}
    \includegraphics[width=0.49\linewidth]{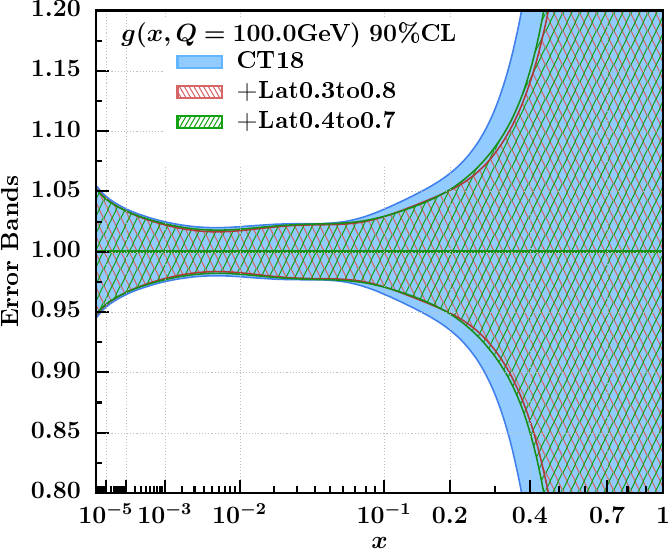}
    \includegraphics[width=0.49\linewidth]{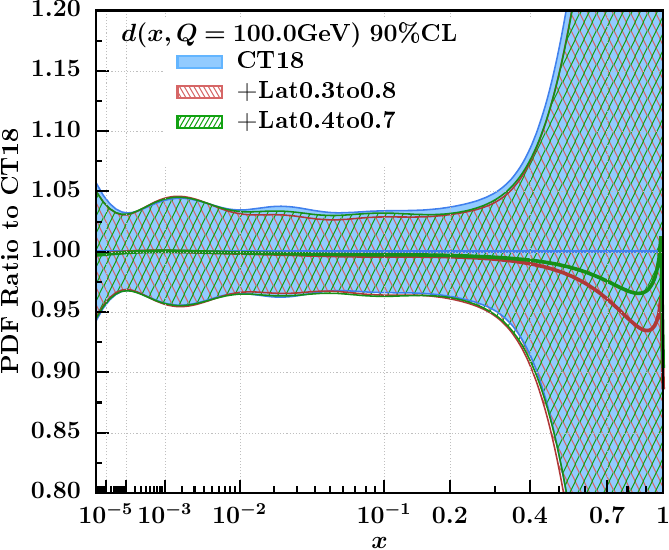}
    \includegraphics[width=0.49\linewidth]{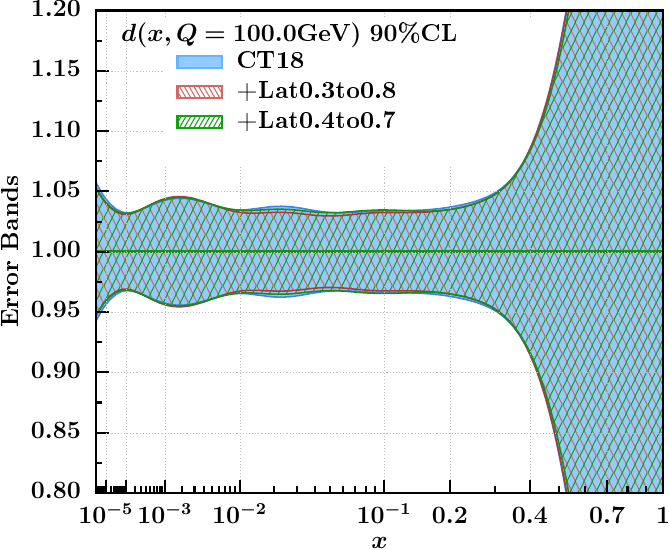}
    \caption{Comparison of the central values (left) and error bands (right) of gluon (upper), down quark (lower) PDFs at $Q=100$~GeV from CT18, CT18+Lat0.3to0.8, and CT18+Lat0.4to0.7 fits.
The error bands are displayed at the 90\% CL.}
    \label{Fig:Q2GeV}
\end{figure}

\begin{figure}[!h]
    \centering
    \includegraphics[width=0.49\linewidth]{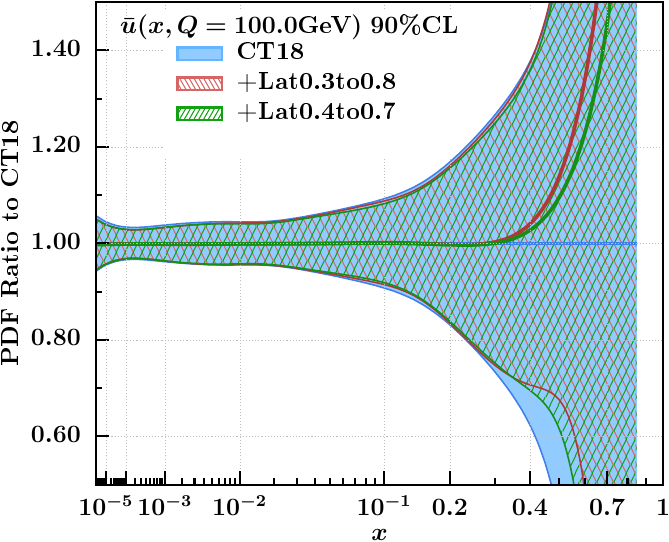}
    \includegraphics[width=0.49\linewidth]{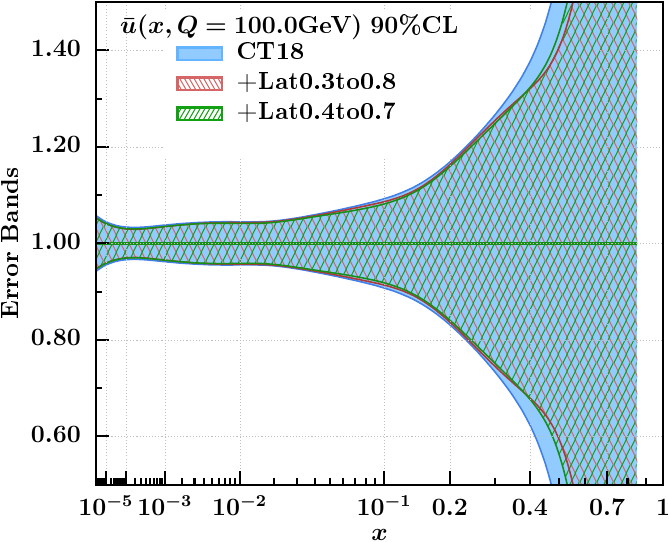}
    \includegraphics[width=0.49\linewidth]{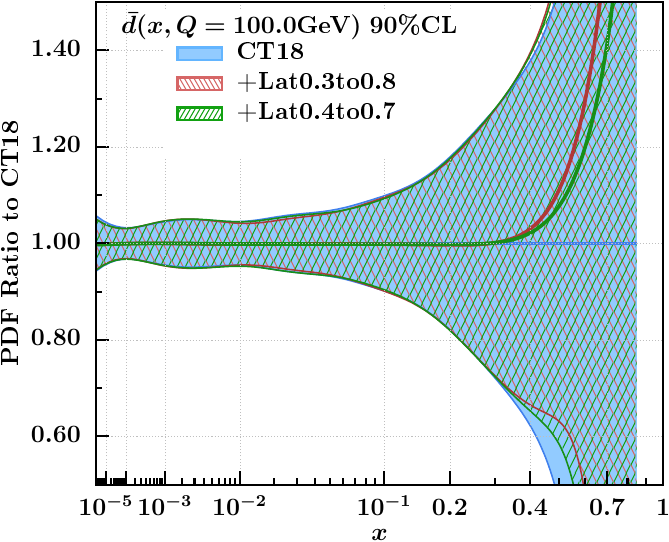}
    \includegraphics[width=0.49\linewidth]{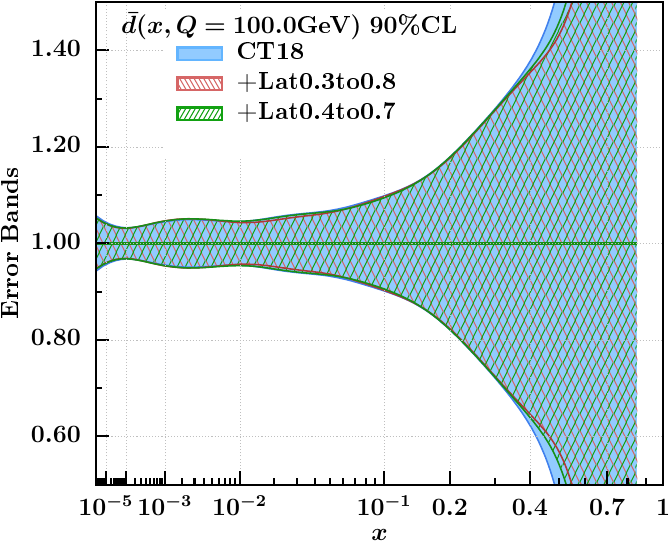}
\includegraphics[width=0.49\linewidth]{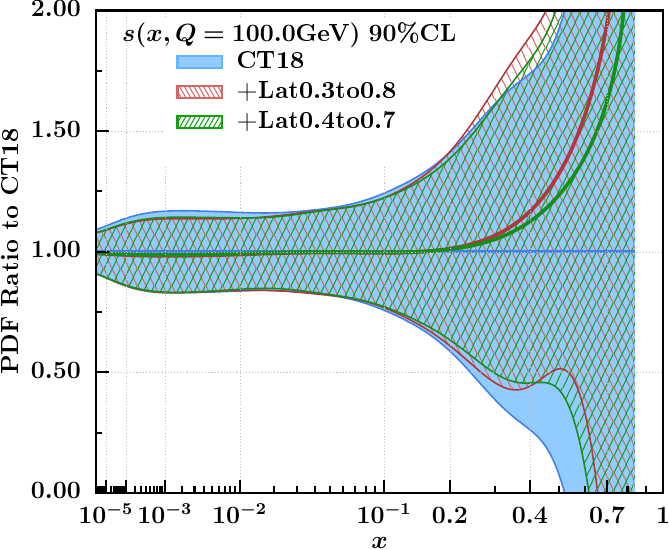}
    \includegraphics[width=0.49\linewidth]{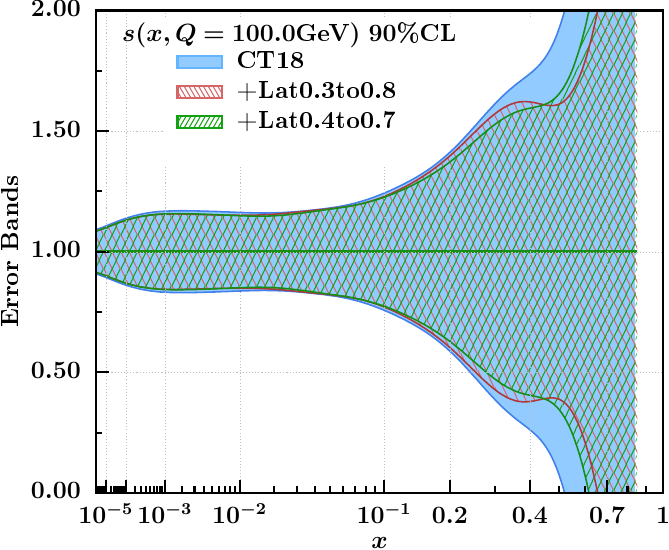}
    \caption{Similar to Fig.~\ref{Fig:Q2GeV} but for
    $\bar u$, ${\bar d}$, and $s$ PDFs.}
    \label{Fig:Q100GeV-ub-db-s}
\end{figure}

In Figs.~\ref{Fig:Q2GeV} and \ref{Fig:Q100GeV-ub-db-s}, we compare the CT18 PDFs at $Q=100$~GeV with the newly obtained PDFs that incorporate lattice gluon datasets in the ranges $x \in [0.3, 0.8]$ and $x \in [0.4, 0.7]$ using Lagrange-multiplier method~\cite{Stump:2001gu}, denoted as ``CT18+Lat0.3to0.8'' and ``CT18+Lat0.4to0.7'', respectively.
We observe that, compared to the CT18 baseline, the inclusion of the lattice gluon dataset leads to a significantly harder large-$x$ gluon PDF while also substantially reducing the associated uncertainty band.
In contrast, the intermediate-$x$ gluon PDF $(10^{-2} \lesssim x \lesssim 10^{-1})$ becomes slightly softer.
Additionally, we find that the lattice gluon dataset slightly reduces the down-quark PDF and its uncertainty while enhancing the $s$ ($\bar{s}$), $\bar{u}$, and $\bar{d}$ sea-quark distributions in the large-$x$ region, again with a notable reduction in uncertainties.
This impact is well understood: sea quarks co-evolve with the gluon.
In contrast, the changes in the down-quark PDF and its corresponding error band are much milder, indicating that valence quarks are already well constrained by existing data, such as DIS.
Between the two choices of momentum fraction ranges $x \in [0.3,0.8]$ and $x \in [0.4,0.7]$ for the lattice gluon PDF inputs, the overall trend remains consistent, though the impact of Lat0.3to0.8 is slightly large.

\begin{figure}[!h]
    \centering
    \includegraphics[width=0.49\linewidth]{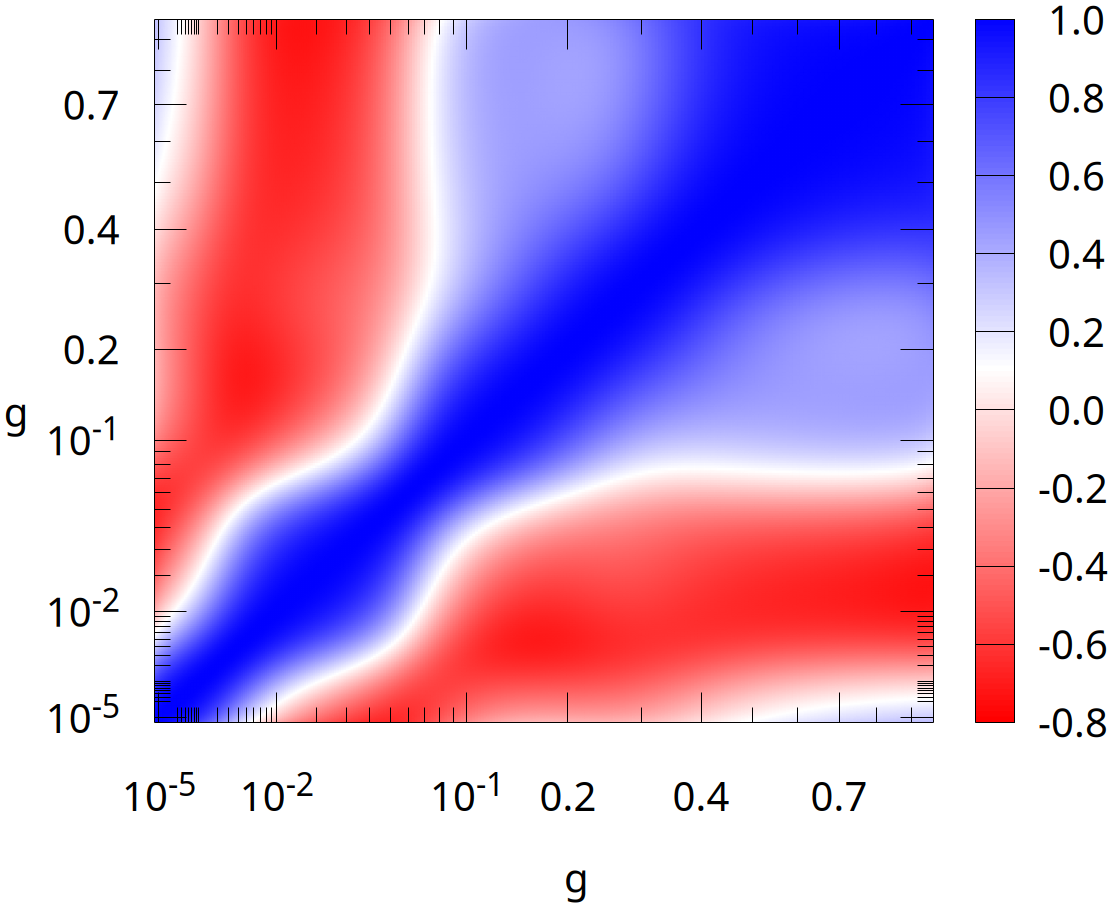}
    \includegraphics[width=0.49\linewidth]{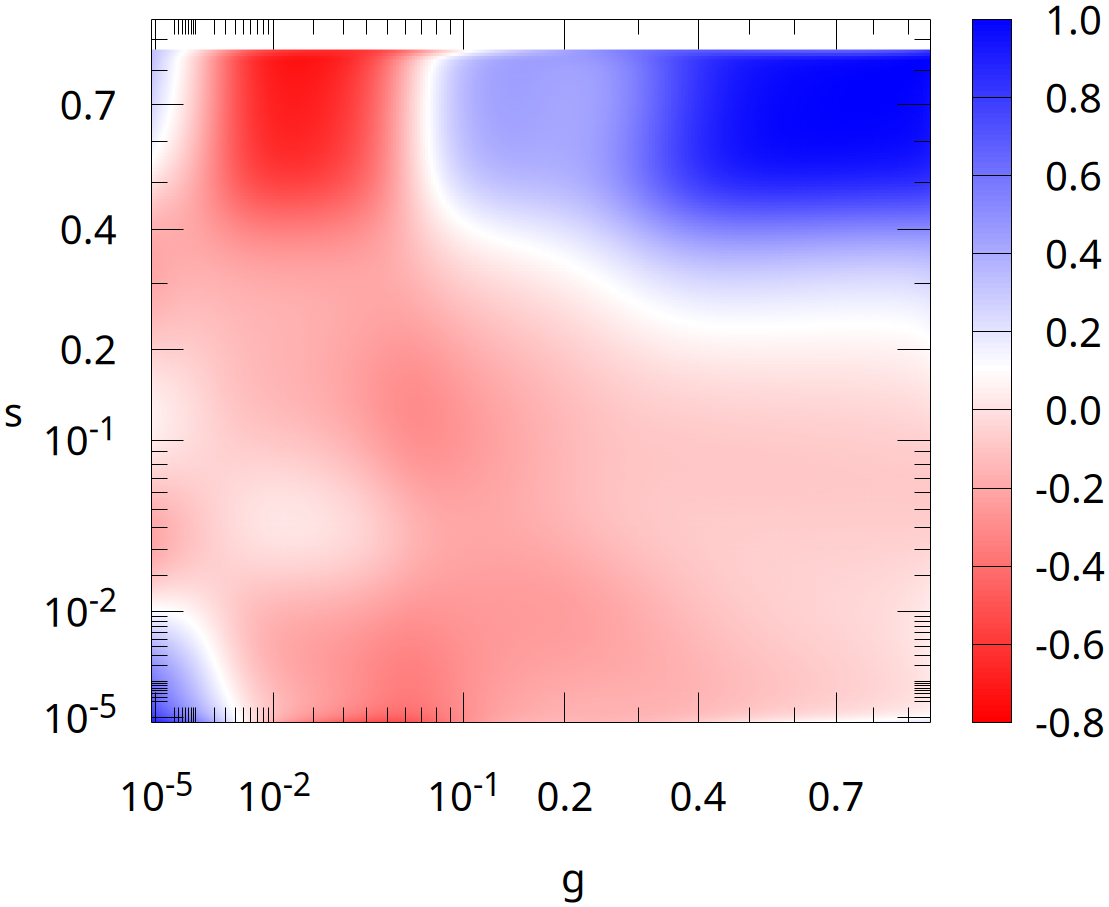}
    \includegraphics[width=0.49\linewidth]{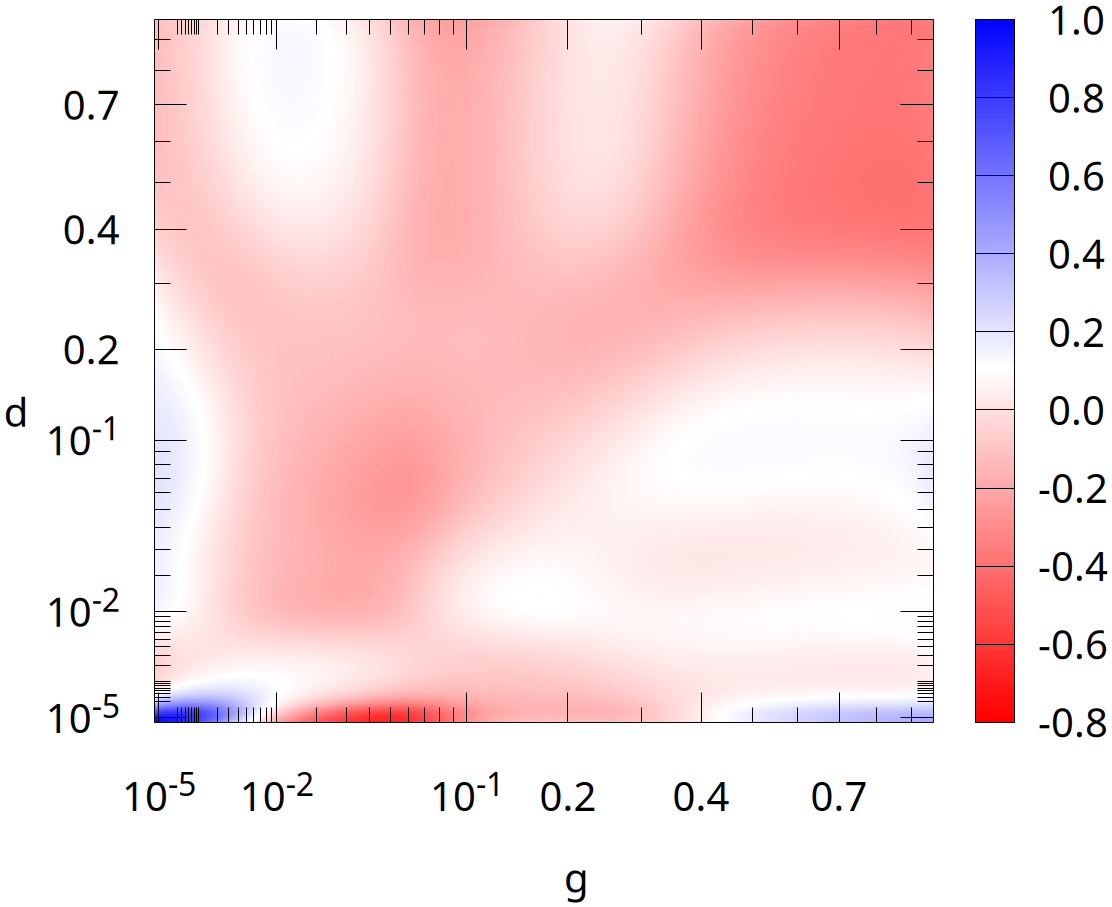}
    \includegraphics[width=0.49\linewidth]{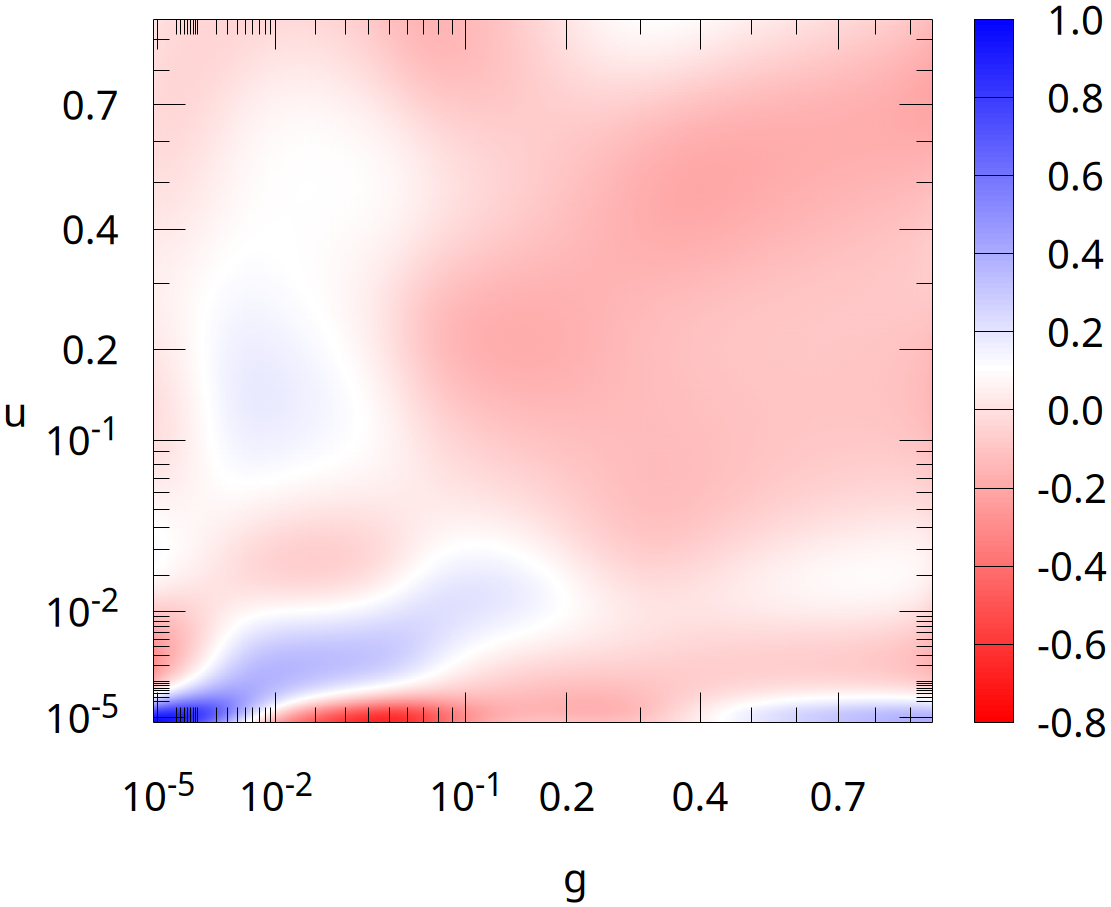}
    \includegraphics[width=0.49\linewidth]{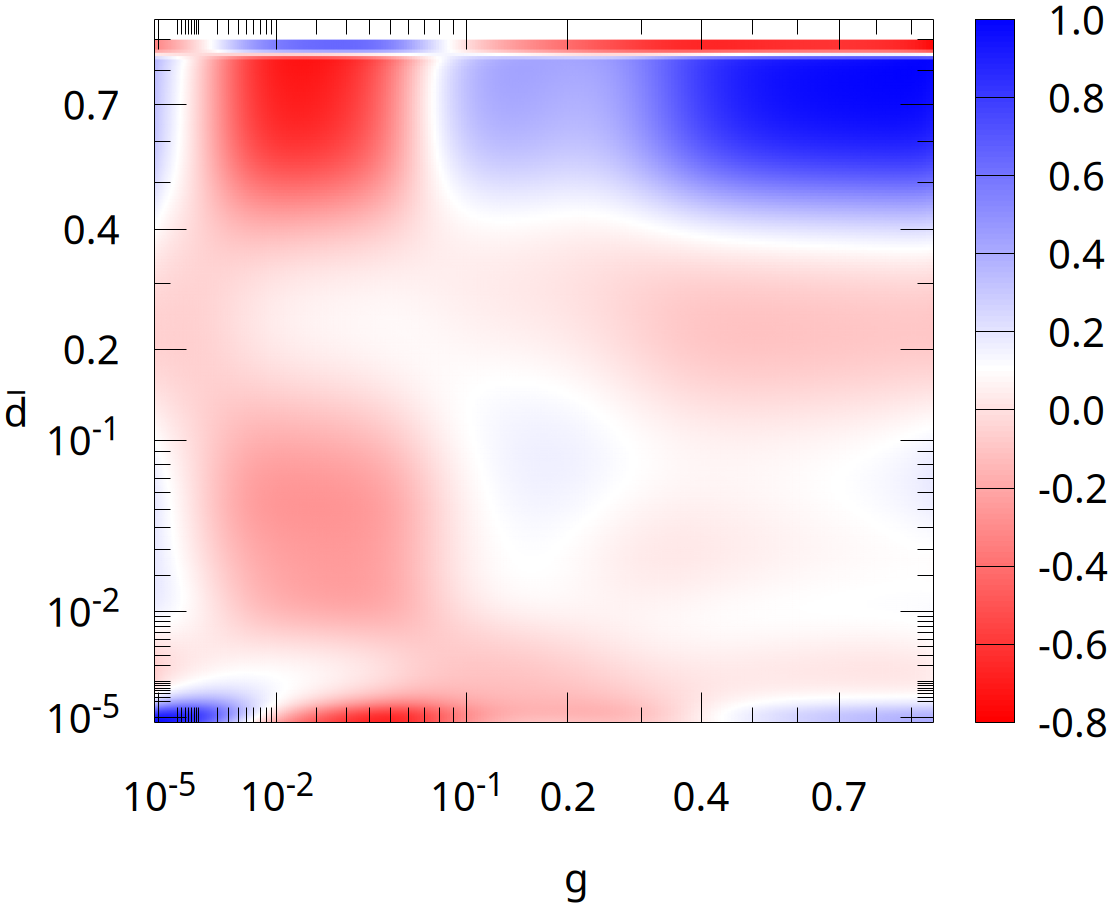}
    \includegraphics[width=0.49\linewidth]{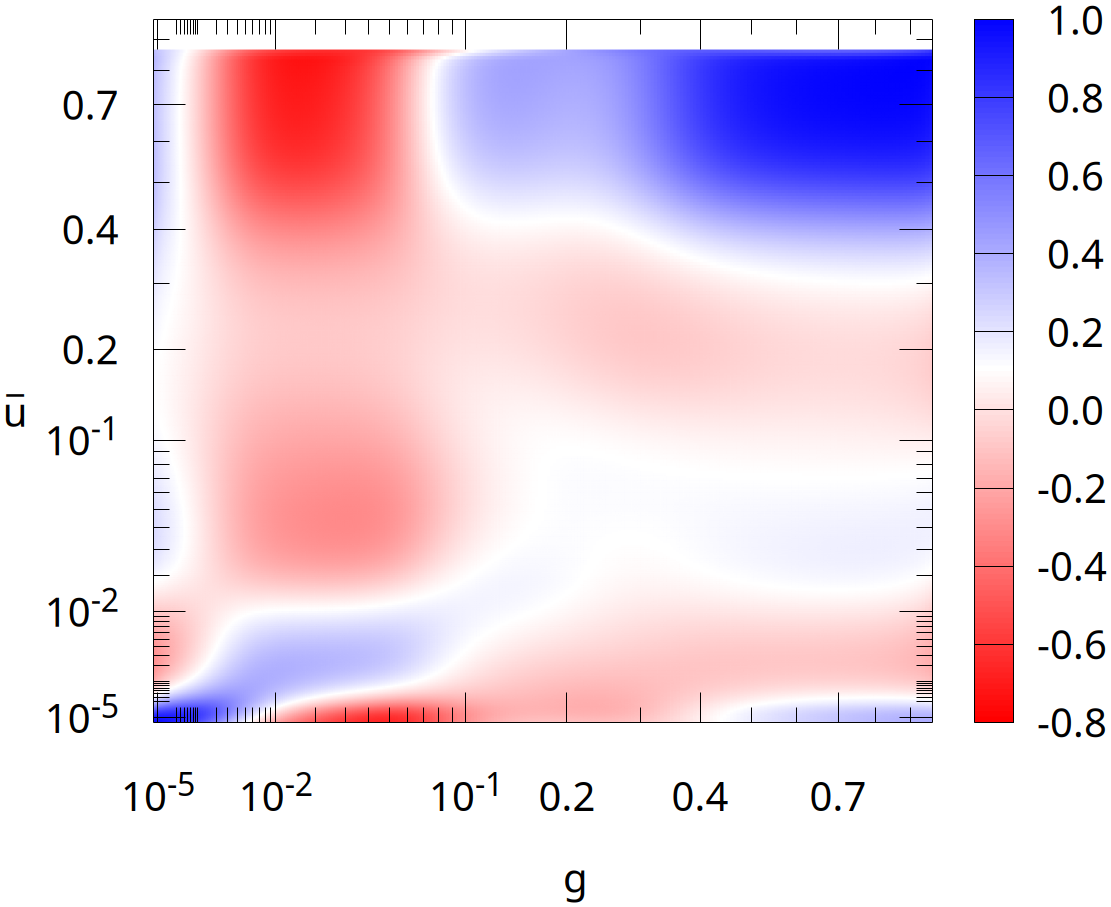}
    \caption{Correlation cosine $\cos\phi$ for the parton pairs $(g,g)$, $(g,s)$, $(g,d)$, $(g,u)$, $(g,\bar{d})$ and $(g,\bar{u})$ at the scale $Q=100$~GeV.}
\label{Fig:corgs}
\end{figure}

To further understand how the gluon inputs from lattice-QCD calculations in the large-$x$ regions ($x \in [0.3, 0.8]$ or $x \in [0.4, 0.7]$) impact the gluon distribution in the mid- and small-$x$ regions and also affect other parton flavors, we performed a correlation analysis of the $g$-$g$, $g$-$s (\bar{s})$, $g$-$d$, $g$-$u$, $g$-$\bar{u}$, and $g$-$\bar{d}$ PDFs at $Q = 100$~GeV, as illustrated in Fig.~\ref{Fig:corgs}.
The results reveal a strong anti-correlation between the gluon PDF in the large-$x$ region and the gluon PDF in the mid- and small-$x$ regions.
Consequently, the lattice dataset suggests a harder gluon distribution in the large-$x$ region, resulting in an upward shift in the gluon PDF.
This shift, in turn, decreases the gluon PDF in the mid- and small-$x$ regions due to the strong anti-correlation.
We also observed a strong correlation between the gluon and the $s(\bar{s})$, $\bar{u}$, and $\bar{d}$ PDFs in the large-$x$ regions, which contributes to an increase in the $s(\bar{s})$, $\bar{u}$, and $\bar{d}$ PDFs in that region.

\begin{figure}[!h]
    \centering
    \includegraphics[width=0.9\linewidth]{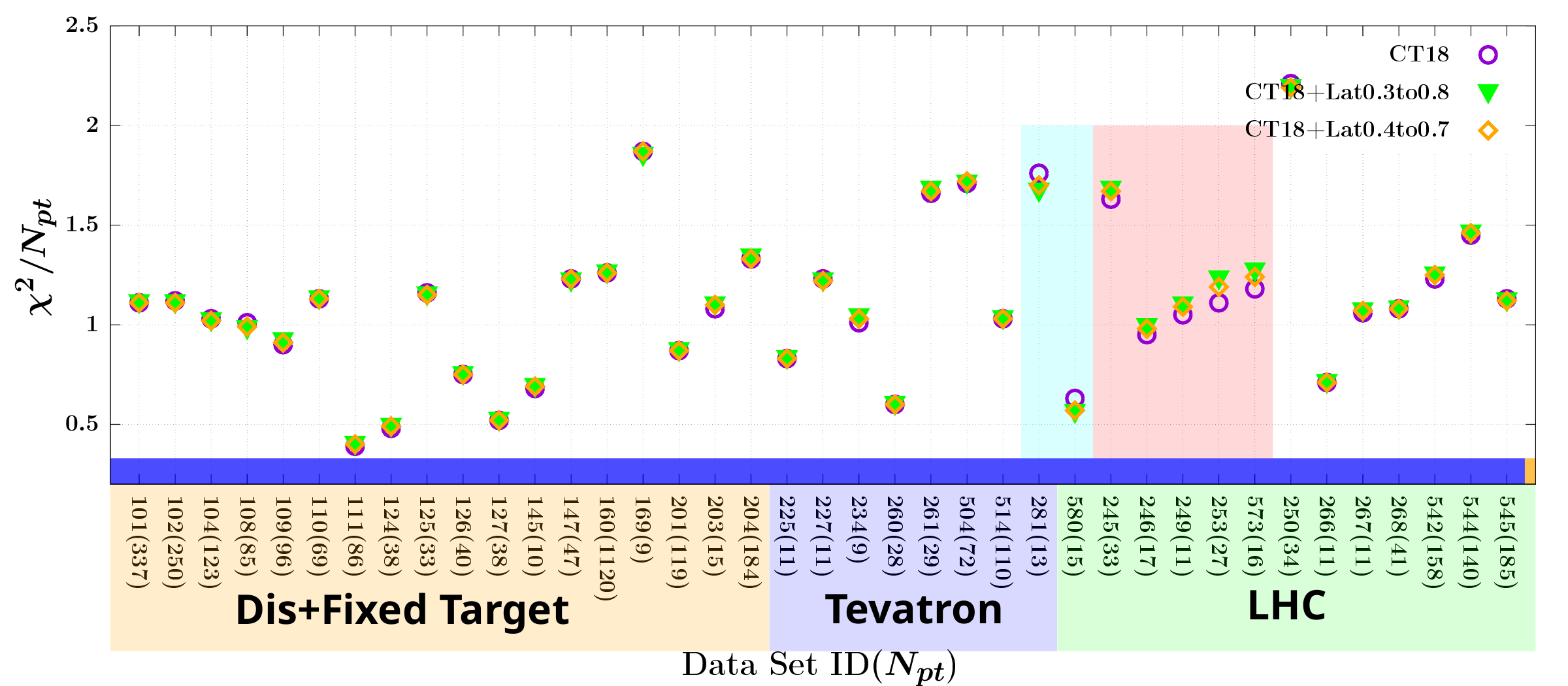}
    \caption{Comparison of $\chi^2/N_\text{pt}$ values for datasets in the CT18, CT18+Lat0.3to0.8 and CT18+Lat0.4to0.7 fits.
The ID~($N_\text{pt}$) of each dataset can be found in Refs.~\cite{Hou:2019efy,Hou:2022onq,Ablat:2024uvg}.}
    \label{Fig:chiCT18data}
\end{figure}

\begin{figure}[!h]
    \centering
    \includegraphics[width=0.8\linewidth]{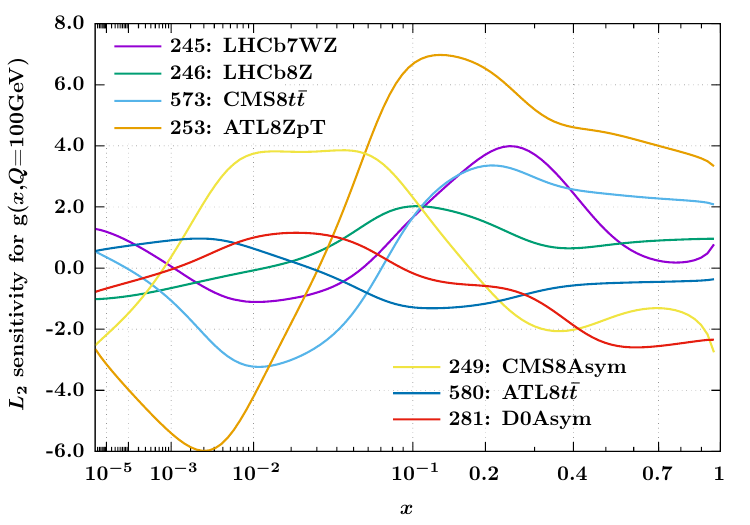}
    \caption{$L_2$ sensitivity plot for the gluon PDF at $Q=100$~GeV in the CT18+Lat0.3to0.8 fit.}
    \label{Fig:L2Sen}
\end{figure}

To assess the goodness-of-fit, we compare the $\chi^2/N_\text{pt}$ values for the datasets in the CT18 fit with those obtained after including the lattice gluon dataset, as shown in Fig.~\ref{Fig:chiCT18data}.
We confirm that the fit quality for the lattice gluon datasets is good for both CT18+Lat0.3to0.8 and CT18+Lat0.4to0.7 PDFs.
In Fig.~\ref{Fig:chiCT18data}, we observe that the overall $\chi^2/N_\text{pt}$ for most datasets remains unchanged after incorporating the lattice gluon dataset, with only a few exhibiting minor variations.
We categorize these small changes into two groups: datasets with an increased $\chi^2$ are marked in red, while those with a decreased $\chi^2$ are marked in cyan.
More specifically, the datasets that show a decrease in $\chi^2$ include only ID~281 (D0Asym, D\O~Run-2 electron charge asymmetry~\cite{D0:2014kma}) and ID~580 (ATL8$t\bar{t}$, ATLAS 8-TeV $t\bar{t}$ production~\cite{ATLAS:2015lsn}).
This suggests that these two datasets exhibit a similar preference for the large-$x$ gluon as the lattice dataset, which we confirm through the $L_2$ sensitivity analysis shown in Fig.~\ref{Fig:L2Sen}.
The $L_2$ sensitivity is defined as the cosine of the correlation angle between $\chi^2$ and the PDFs, weighted by the corresponding Hessian variation $\Delta\chi^2$~\cite{Hobbs:2019gob,Jing:2023isu}.
This quantifies the statistical pulls exerted by individual data and highlights potential tensions among them.
 As seen in Fig.~\ref{Fig:L2Sen}, the large-$x$ gluon PDF shows negative $L_2$ sensitivities for 281 D0Asym and 580 ATL8$t\bar{t}$, indicating a $\chi^2$ reduction trend
and a preference for a hard gluon PDF at large-$x$.
In comparison, five datasets exhibit a minor increase in $\chi^2$, as indicated in red in Fig.~\ref{Fig:chiCT18data}:
ID~245 LHCb7WZ (LHCb 7-TeV $W/Z$ forward rapidity~\cite{LHCb:2015okr}),
ID~246 LHCb8Z (LHCb 8-TeV $Z\rightarrow e^{-} e^{+}$ forward rapidity~\cite{LHCb:2015kwa}),
ID~253 ATL8ZpT (ATLAS 8-TeV $Z$ $p_T$~\cite{ATLAS:2015iiu}),
ID~249 CMS8Asym (CMS 8-TeV $\mu$ charge asymmetry~\cite{CMS:2016qqr}),
and 573 CMS8$t\bar{t}$ (CMS 8-TeV $t\bar{t}$ production~\cite{CMS:2017iqf}).
Similarly, we present the corresponding $L_2$ sensitivity in Fig.~\ref{Fig:L2Sen}, which further confirms the observed $\chi^2$ variations.
It is important to note that although the CMS and ATLAS 8-TeV $t\bar{t}$ datasets both probe the large-$x$ gluon PDF, a small tension between these two datasets, as discussed in
Ref.~\cite{Hou:2019efy}, leads to the different behaviors observed for the $\chi^2$ change in Fig.~\ref{Fig:chiCT18data} and the $L_2$ sensitivity in Fig.~\ref{Fig:L2Sen}.

\begin{figure}[!h]
    \centering
    \includegraphics[width=0.49\linewidth]{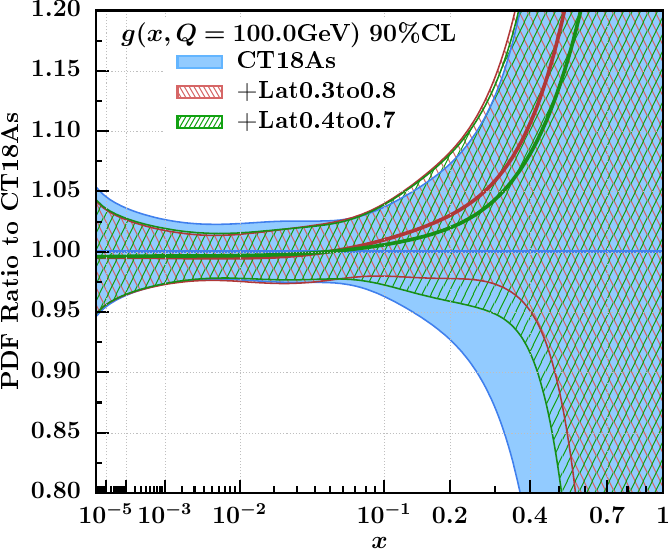}
    \includegraphics[width=0.49\linewidth]{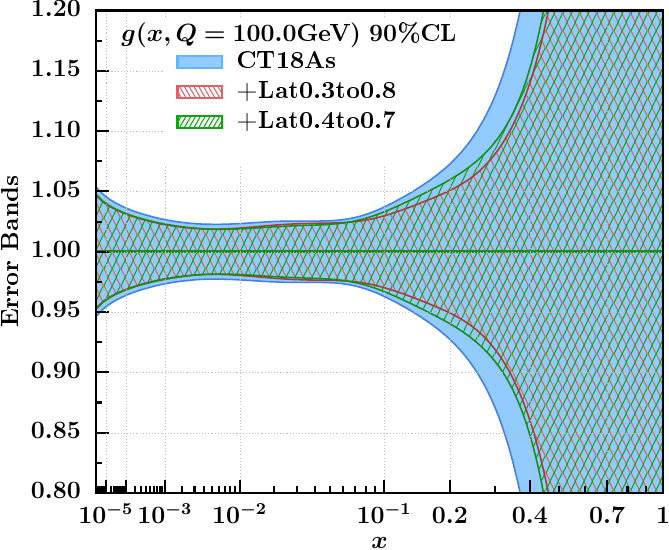}
    \includegraphics[width=0.49\linewidth]{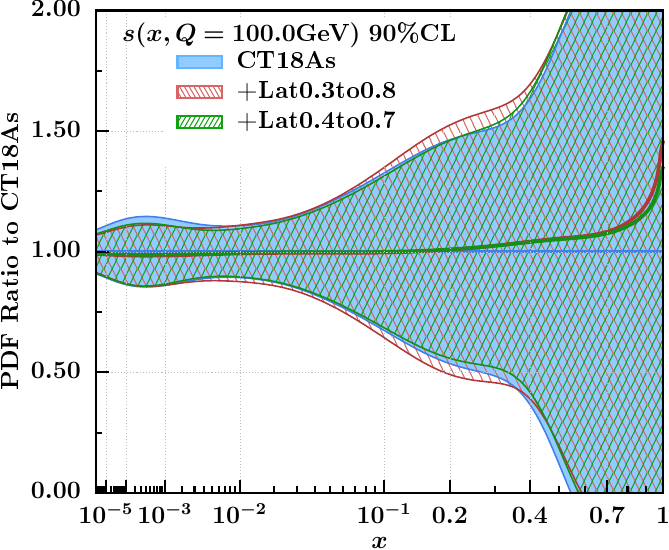}
    \includegraphics[width=0.49\linewidth]{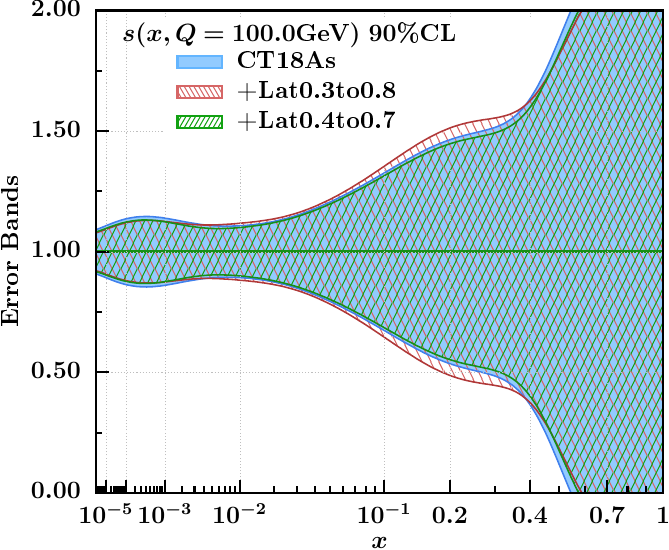}
    \caption{The comparisons of the gluon and strange PDFs from CT18As, CT18As+Lat0.3to0.8, and CT18As+Lat0.4to0.7.
    }
    \label{fig:CT18AsLatGluon}
\end{figure}

Another important question we seek to address is the extent to which the reduction in PDF uncertainty depends on the PDF parameterization, particularly for sea quarks, such as the strange PDF shown in Fig.~\ref{Fig:Q100GeV-ub-db-s}.
Therefore, we conducted a parallel study by incorporating the lattice dataset into the global fit based on the CT18As parameterization~\cite{Hou:2022onq}.
In the CT18As fit, we include the additional ATLAS 7-TeV $W, Z$ precision dataset~\cite{ATLAS:2016nqi} and introduce two additional PDF shape parameters to allow for a nonzero strangeness asymmetry, $s_{-}(x) = s(x) - \bar{s}(x)$, at the starting scale $Q_0$.
The global-fit results for the gluon and strange-quark PDFs with the lattice gluon inputs are presented in Fig.~\ref{fig:CT18AsLatGluon}.
This figure shows that the impact of the lattice dataset on the gluon PDF is largely consistent with the CT18 case in Fig.~\ref{Fig:Q2GeV}, both in terms of the central value and the error band.
However, when examining the strange PDF, we observe a slight reduction in the central value at large-$x$, mainly driven by the more flexible parameterizations of the $s$ and $\bar{s}$ PDFs.
A notable difference from the CT18 scenario is that the uncertainties of the $s$- and $\bar{s}$ PDFs largely remain unchanged with the inclusion of the lattice dataset.
This can be attributed to the weakened correlation between the $s$, $\bar{s}$, and $g$ PDFs at large-$x$, due to the additional flexibility in the CT18As framework.
In other words, the indirect impact of lattice gluon dataset on sea quarks, such as the $s$- and $\bar{s}$ PDFs, can be parameterization dependent, as it originates from an indirect propagation in co-evolution and the momentum sum rule.
Nevertheless, we confirm that the effects of the lattice gluon PDF dataset on the $\bar{u}$- and $\bar{d}$ PDFs are quite similar between the fits with the CT18As and the CT18 baseline.
Furthermore, the impact of lattice input on the gluon PDF is largely parameterization independent, which remains the primary focus of this work.

\subsection{Interplay with the collider jet datasets}
\label{sec:interplay}

\begin{figure}[!h]
    \centering
    \includegraphics[width=0.49\linewidth]{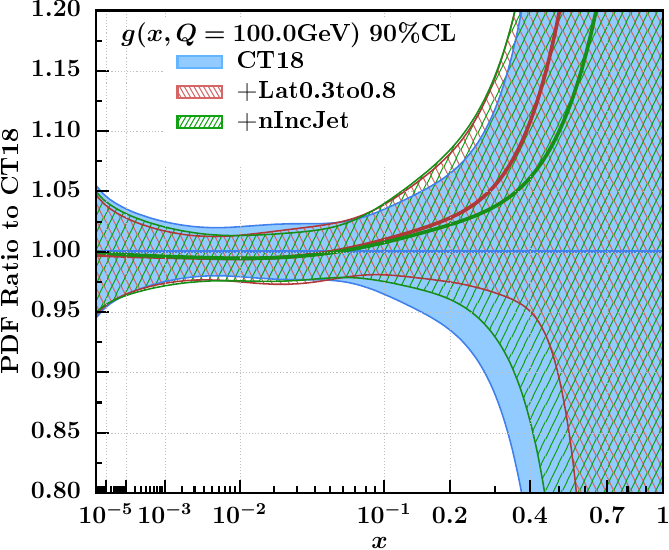}
    \includegraphics[width=0.49\linewidth]{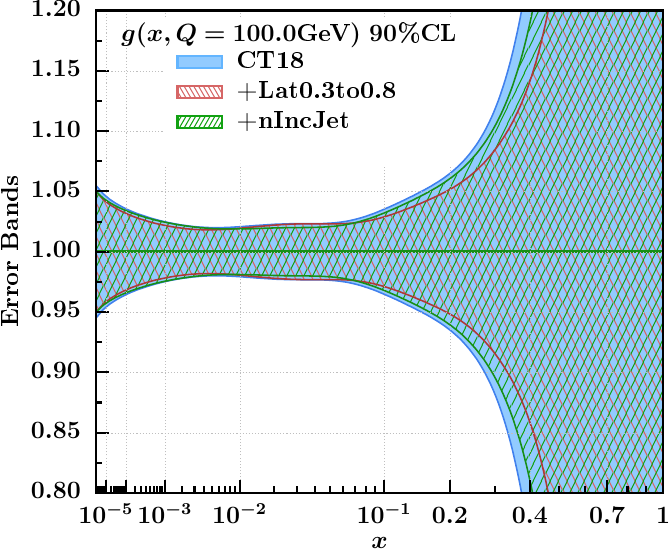}
    \caption{Impact of lattice gluon and post-CT18 inclusive jet datasets on the CT18 gluon PDF.}
    \label{fig:Latt2jet}
\end{figure}

In a modern global analysis, such as CT18~\cite{Hou:2019efy}, the large-$x$ gluon PDF is primarily constrained by jet-production measurements.
Therefore, we investigate the interplay between inclusive-jet and lattice-gluon datasets.
In Fig.~\ref{fig:Latt2jet}, we first compare the impact of the lattice gluon dataset (Latt0.3to0.8) with that of the latest post-CT18 ATLAS 8-TeV~\cite{ATLAS:2017kux}, 13-TeV~\cite{ATLAS:2017ble} and CMS 13-TeV~\cite{CMS:2021yzl} inclusive-jet datasets (nIncJet), examined in our recent study~\cite{Ablat:2024uvg}.
Interestingly, we find that the lattice input shifts the large-$x$ gluon distribution in the same direction as the latest inclusive-jet datasets, with the lattice dataset exerting an even stronger pull.
When examining the PDF uncertainty band in Fig.~\ref{fig:Latt2jet} (right), we reach a similar conclusion: both the lattice input and the inclusive jet datasets contribute to reducing gluon PDF uncertainty, with the lattice dataset leading to a slightly greater reduction in the error band.
As a side note, we observe that both the lattice input and inclusive jet datasets slightly suppress the intermediate-$x$ ($10^{-2} \sim 10^{-1}$) gluon PDF, counterbalancing the pull at large-$x$.

\begin{figure}[!h]
    \centering
    \includegraphics[width=0.49\linewidth]{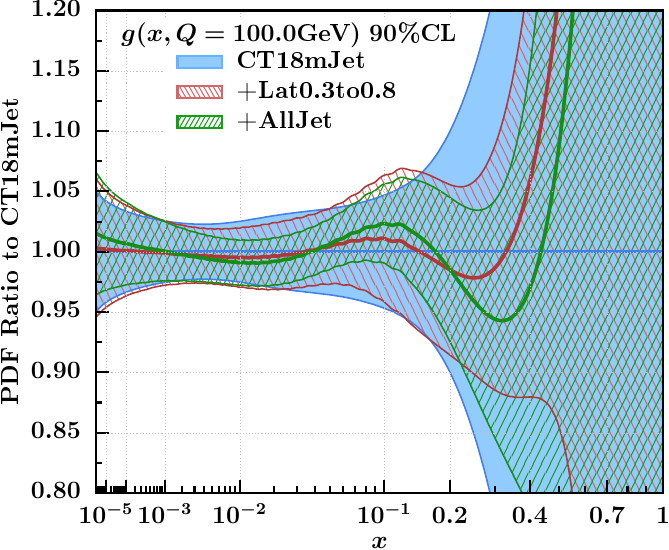}
    \includegraphics[width=0.49\linewidth]{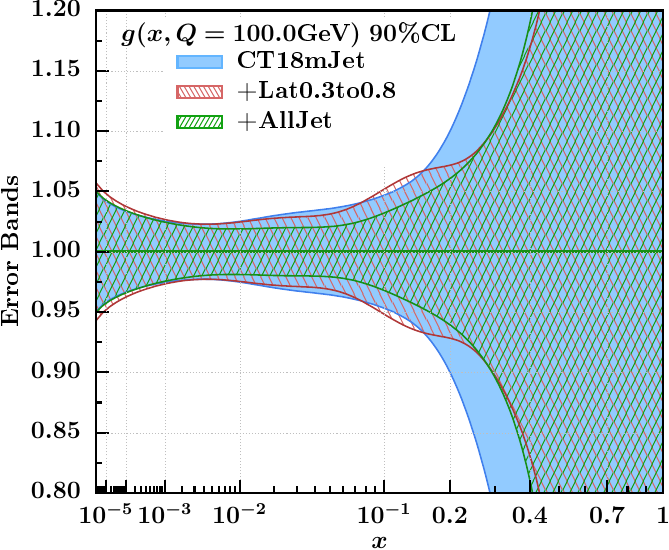}
\caption{Similar to Fig.~\ref{fig:Latt2jet}, but with CT18mJet baseline.}
    \label{Fig:CT18mJet}
\end{figure}

To further investigate the impact of the lattice gluon dataset on the gluon PDF, we performed an additional fit based on the CT18 non-jet datasets.
Specifically, we excluded all inclusive jet datasets from the Tevatron~\cite{CDF:2008hmn,D0:2008nou} and LHC~\cite{CMS:2014nvq,ATLAS:2014riz,CMS:2016lna} experiments, comprising approximately 1,000 data points, collectively referred to as the "AllJet" from the CT18 fit.
We designated this modified fit as CT18mJet.
As illustrated in Fig. \ref{Fig:CT18mJet}, using CT18mJet as the baseline, incorporating the AllJet datasets, further constrains the gluon PDF in the intermediate and large-$x$ regions, with a preference for a harder gluon PDF at large-$x$.
This result aligns with the CT18+nIncJet fit, as reported in Ref. \cite{Ablat:2024uvg}, in which the impact of each jet dataset from collider experiments was also discussed.
Furthermore, we observe that the Lat0.3to0.8 lattice dataset primarily constrains the CT18mJet gluon PDF at $x > 0.3$.
For $x < 0.3$, the AllJet datasets from Tevatron and LHC impose additional constraints on the gluon PDF, whereas for $x > 0.5$, the lattice dataset provides constraints comparable to, or even stronger than, those from the current AllJet collider datasets.
This observation highlights the unique and complementary role of the lattice dataset in constraining the gluon distribution in the large-$x$ region, where experimental measurements are often sparse or less precise.

\section{Phenomenological Implications}
\label{sec:pheno}

\begin{figure}[!h]
    \centering
    \includegraphics[width=0.49\linewidth]{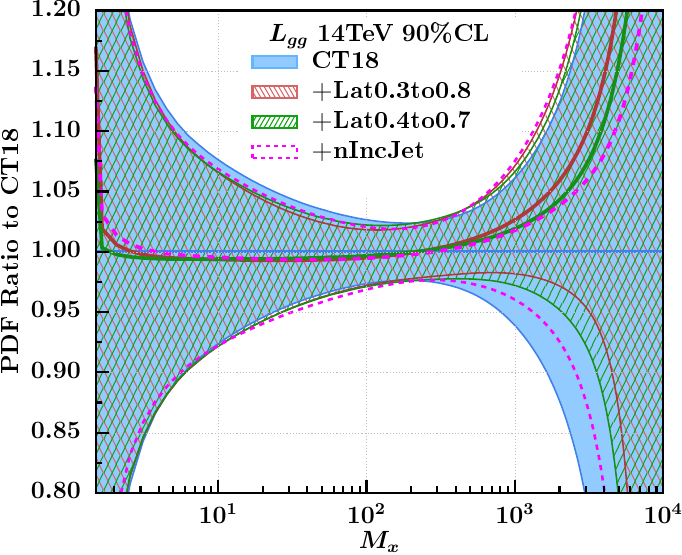}
    \includegraphics[width=0.49\linewidth]{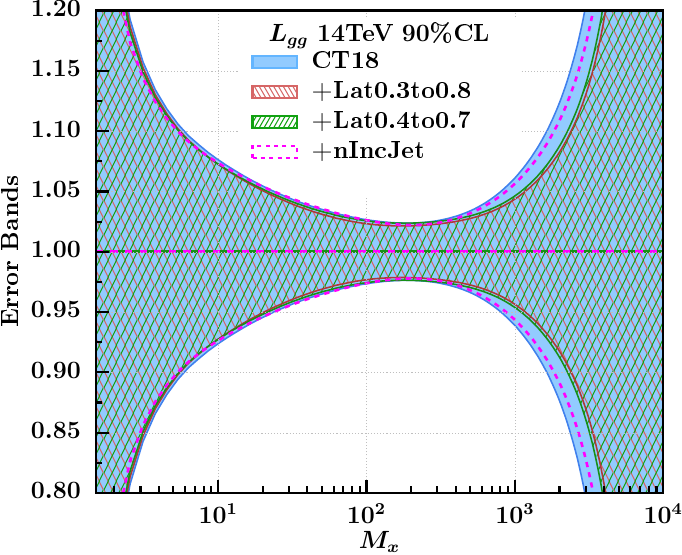}
    \caption{Comparison of the $gg$ parton luminosity (left) and the corresponding uncertainty (right) at LHC 14~TeV for CT18, CT18+Lat0.3to0.8, CT18+Lat0.4to0.7, and CT18+nIncJet fits.}
    \label{Fig:PDFLuminosity}
\end{figure}

In this subsection, we explore the phenomenological implications of the lattice gluon dataset, with a primary focus on the gluon PDF and related processes at the LHC.
Before analyzing a specific physical process, we first compare the gluon-gluon ($gg$) parton luminosity, $L_{gg}$, before and after incorporating the lattice dataset within the CT18 framework, as shown in Fig.~\ref{Fig:PDFLuminosity}.
The partonic luminosity is defined as follows~\cite{Campbell:2006wx}:
\begin{equation}\label{eq:Lgg}
L_{gg}(s,M_X^2)=\frac{1}{s}\int_{\tau=M_X^2/s}^1\frac{\dd x}{x}g(x,Q^2)g(\tau/x,Q^2),
\end{equation}
typical renormalization and factorization scales are chosen as the invariant mass, $Q=M_X$.
We observe that $L_{gg}$ exhibits a noticeable enhancement at large invariant mass $M_X$ after incorporating the lattice dataset, in both the Latt0.3to0.8 and Latt0.4to0.7 scenarios.
For comparison, we also include the fit with the post-CT18 inclusive jet datasets as a reference.
As expected, the impact of the lattice gluon input is stronger, which naturally follows from the gluon PDF shown in Fig.~\ref{fig:Latt2jet}.
When examining the corresponding uncertainty in Fig.~\ref{Fig:PDFLuminosity} (right), we find that the inclusion of the lattice dataset significantly reduces the uncertainty, even compared to the CT18+nJet fit.
This underscores the constraining power of the lattice input.

\begin{figure}[!h]
    \centering
    \includegraphics[width=0.49\linewidth]{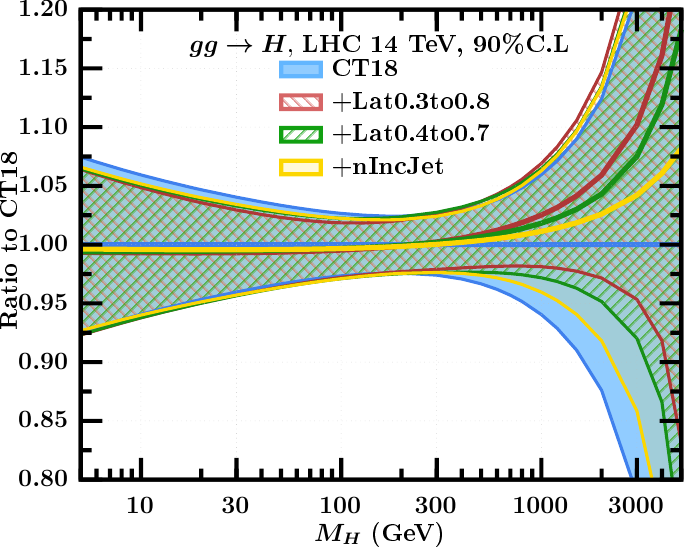}
    \includegraphics[width=0.49\linewidth]{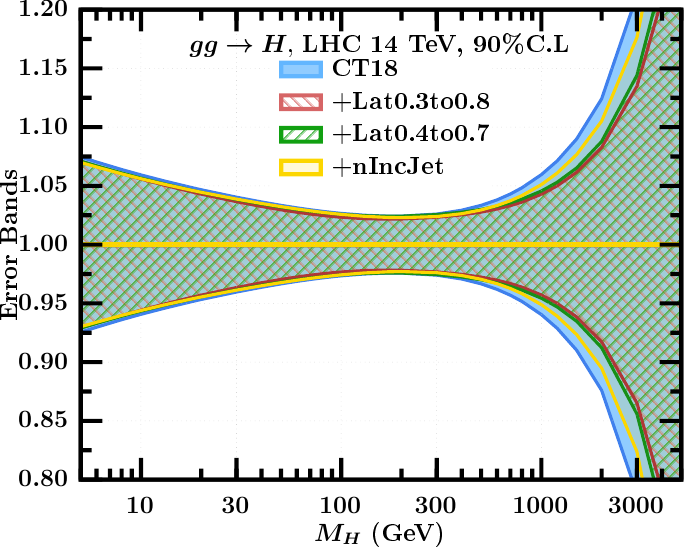}
    \caption{Comparison of the cross section of a Higgs-like scalar with mass $M_H$ production at the LHC 14~TeV, among the CT18, CT18+Lat0.3to0.8, CT18+Lat0.4to0.7, and CT18+nIncJet predictions (left) and the corresponding 90\% CL uncertainty (right).}
    \label{Fig:N3LO-Calc-H}
\end{figure}

As a natural implication of $L_{gg}$, we present the cross section and its corresponding uncertainty for Higgs-like scalar production via gluon-gluon fusion at the LHC with a center-of-mass energy of 14~TeV, as shown in Fig.~\ref{Fig:N3LO-Calc-H}.
The theoretical calculation is performed using the \texttt{n3loxs}~\cite{Baglio:2022wzu} code, with the scalar mass $M_H$ varying from 5~GeV to 5~TeV, matching the partonic invariant mass $M_X$ in Eq.~(\ref{eq:Lgg}).
For a scalar mass equal to the SM Higgs value, $M_H=125$~GeV, we have validated the calculation using \texttt{ggHiggs}~\cite{Bonvini:2016frm}, which shows perfect agreement.
We observe that the trends of $L_{gg}$ in Fig.~\ref{Fig:PDFLuminosity} are reflected in the cross section shown in Fig.~\ref{Fig:N3LO-Calc-H}.
This implies that for a large scalar mass, such as $M_H > 500$~GeV, the inclusion of a lattice gluon dataset leads to a significant enhancement of the cross section.
This effect is stronger than that observed with the latest inclusive jet measurements at the LHC.
Furthermore, the lattice gluon dataset helps reduce the corresponding PDF uncertainty, once again showing a slightly stronger effect than the post-CT18 inclusive jet datasets.
This is consistent with the comparison of luminosity error bands shown in Fig.~\ref{Fig:PDFLuminosity} (right).

In a more realistic scenario, we focus on the production of top-quark ($t\bar{t}$) pairs and their associated production with an additional jet, Higgs, or $Z$ boson at the LHC with a center-of-mass energy of 14~TeV, where the gluon-gluon fusion subprocess dominates.
The top-quark pair cross section is calculated using \texttt{Top++}~\cite{Czakon:2011xx} at NNLO, including soft-gluon resummation up to NNLL.
The factorization and renormalization scales are set to the top-quark pole mass, $m_t = 172.5$~GeV.
Other processes are computed with \texttt{MadGraph\_aMC@NLO}~\cite{Alwall:2014hca,Frederix:2018nkq} at NLO, with the scale set to $M_T/2$, where $M_T$ is the sum of the transverse masses of the final-state particles.
The jet is defined in the fiducial region as
\begin{equation}
p_T^j > 25~\text{GeV},\quad |\eta_j| < 3,\quad \Delta R = 0.7.
\end{equation}

\begin{figure}[!h]
    \centering
    \includegraphics[width=0.49\linewidth]{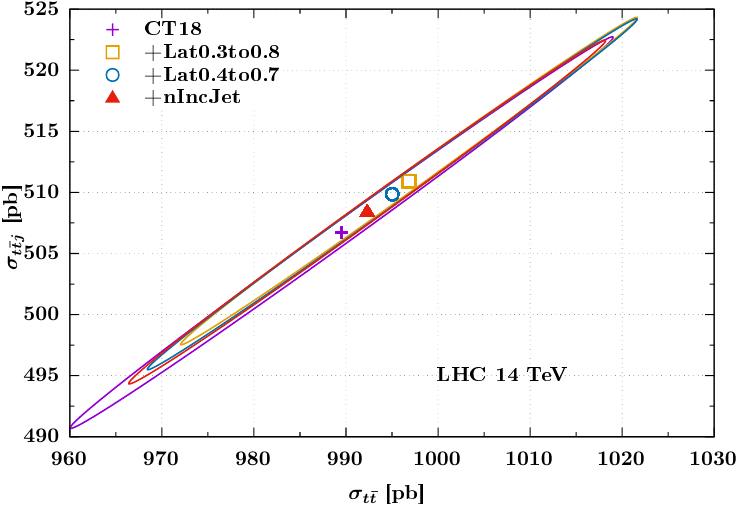}
    \includegraphics[width=0.49\linewidth]{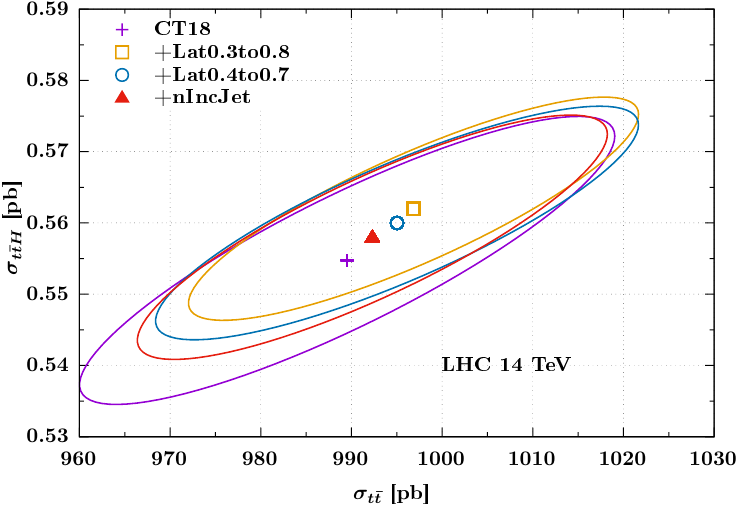}
    \includegraphics[width=0.49\linewidth]{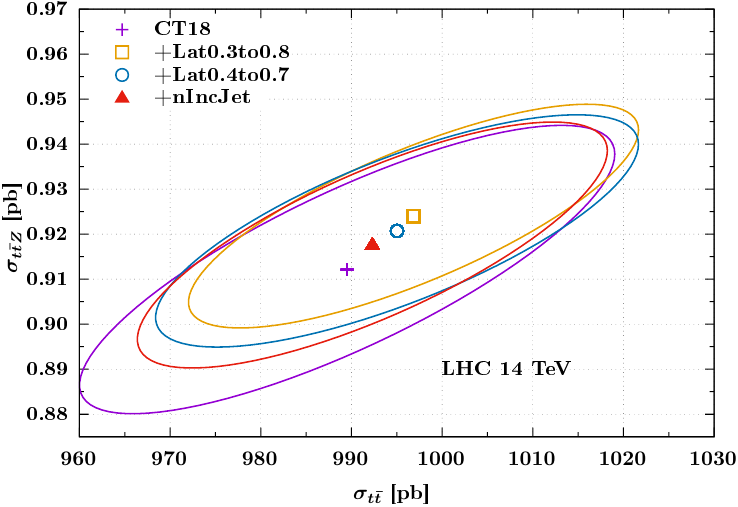}
    \includegraphics[width=0.49\linewidth]{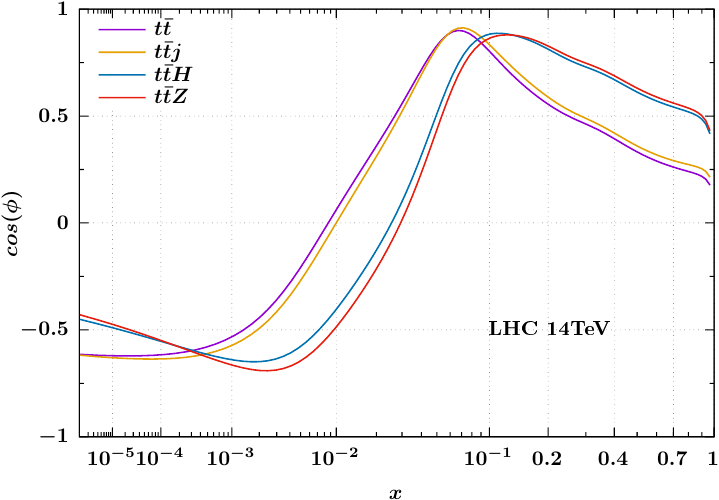}
    \caption{The correlation ellipses at 90\% CL and correlation cosines with the $Q=100$~GeV gluon PDF for the production of $t\bar{t}$, $t\bar{t}j$, $t\bar{t}H$ and $t\bar{t}Z$ at the LHC 14~TeV.}
     \label{Fig:CorrCosine}
\end{figure}

In Fig.~\ref{Fig:CorrCosine}, we present the corresponding correlation ellipses at the 90\% CL along with the correlation cosine with the gluon PDF.
First, we observe a strong positive correlation among all these processes, particularly between $t\bar{t}$ and $t\bar{t}j$, which naturally follows from the dominance of gluon-gluon fusion.
Additionally, we find that all cross sections are enhanced by the inclusion of the lattice dataset, with a larger impact compared to the post-CT18 inclusive jet datasets.
This is consistent with the comparison of the gluon PDF shown in Fig.~\ref{fig:Latt2jet} and the $gg$ parton luminosity presented in Fig.~\ref{Fig:PDFLuminosity}.
Furthermore, we provide the corresponding correlation cosine with the gluon PDF in Fig.~\ref{Fig:CorrCosine} (lower right) as additional confirmation.
We observe a reduction in the associated uncertainties, which stems from the decreased uncertainty in the large-$x$ gluon PDF and the high-invariant-mass $gg$ luminosity.

\section{From Lattice Matrix elements to Global PDF Constraints} 
\label{Sec:Analysis2}

In the previous sections, the underlying philosophy was to analyze the lattice data independently, producing an $x$-dependent PDF with associated uncertainties that can be used directly as theory input to a global fit.
In contrast, the pseudo-PDF approach is based on a factorization theorem that relates the PDF to the lattice matrix elements, namely the RpITDs shown on the left-hand side of Fig.~\ref{fig:LatGluon}.
This formulation introduces an inverse problem, analogous to that encountered when extracting PDFs from experimental cross sections. 
This motivates an alternative strategy to incorporate the lattice matrix elements directly into the global fit, treating them in the same manner as experimental data. 
In principle, the information contained in the matrix elements and in the resulting lattice PDF is identical, differing only in its representation. 
In the limit of sufficiently small statistical and systematic uncertainties, both approaches should yield consistent results. 
Motivated by this expectation, we investigate this alternative method and compare its results with those obtained from the PDF-based approach.

Stating the pseudo-PDF factorization theorem explicitly, the RpITDs are connected to the light cone PDF via the relation~\cite{Balitsky:2019krf, Balitsky:2021qsr, MorrisIII:2022fav} 
\begin{eqnarray}
     && \hspace*{-0.5cm}
\mathcal{M}(\nu, z^2)  
= \int_0^1 d x \frac{xg(x,\mu^2)}{\langle x_g \rangle}
  \left[R_{gg}(x\nu, \mu^2z^2) 
      + R_r(x\nu, \mu^2z^2) \frac{\langle x_\Sigma \rangle}{\langle x_g \rangle} \right]  
      \notag\\
     && \hspace*{-0.5cm}
+ \int_0^1 d x \frac{x\Sigma(x,\mu^2)}{\langle x_g \rangle} R_{gq}(x\nu, \mu^2z^2) 
+ \mathcal{O}(m^2z^2, z^2 \Lambda_\text{QCD}),
\label{eq:matching}
\end{eqnarray}
where $\mathcal{M}(\nu, z^2)$ is the RpITD, $\nu=zP_z$ is the Ioffe time for the matrix element separation $z$ and boost momentum $P_z$, $\mu^2$ is the $\overline{\rm MS}$ scheme renormalization scale, $\Sigma$ denotes the quark singlet, and $\langle x_f\rangle = \int_0^1 dx \,xf(x, \mu^2)$ are the momentum fractions.
$R_{gg}$ is the gluon-gluon matching kernel, while $R_{gq}$ explicitly handles the gluon-quark mixing and $R_r$ is necessary to proper normalization when quark mixing is included.
The effects of the quark mixing are very small compared to current statistical errors and other systematics but are included for thoroughness.  
The $\mathcal{O}(z^2 m^2, z^2 \Lambda_{\text{QCD}}^2)$ term describes the power corrections, arising from the finite spatial separation in the Euclidean correlators.
These correction vanish as the separation approaches the light-like limit, $z^2 \rightarrow 0$.
The lattice RpITD is normalized to $\mathcal{M}(\nu=0,z^2)=1$ in the double ratio renormalization scheme applied to the data.
The factors of $\langle x_g\rangle$ and $\langle x_\Sigma\rangle$ on the right side of Eq.~(\ref{eq:matching}) emerge as a consequence of this scheme choice, and the normalization holds only if the momentum sum rule $\langle x_g \rangle+\langle x_\Sigma \rangle=1$ is satisfied.
The most convenient and informative way to include the RpITD data into a phenomenological fit is to keep all terms on the right phenomenological and perturbative, rather than attempting to include lattice information about $\langle x_g \rangle$ or $\langle x_\Sigma \rangle$ into the matching relation.

\begin{figure}
    \centering
    \includegraphics[width=0.45\linewidth]{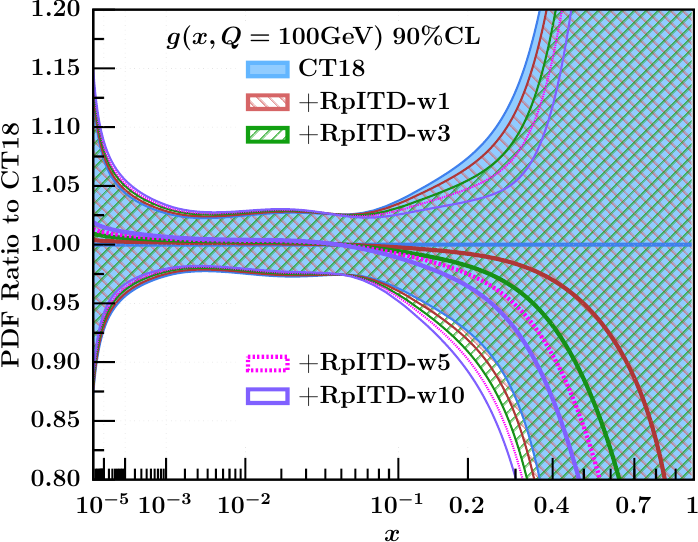}
    \includegraphics[width=0.45\linewidth]{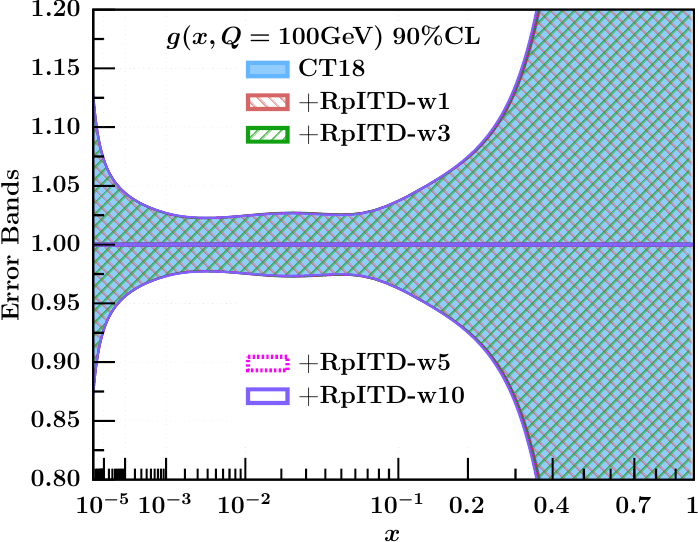}
    \caption{Comparison of the central values (left) and error bands (right) of the gluon PDFs at $Q = 100$ GeV from CT18 and CT18+RpITD at various weights ranging from 1 to 10.}
    \label{fig:ePump_gluons}
\end{figure}

We may calculate theory RpITDs from the CT18 PDF set using Eq.~\eqref{eq:matching} and compare with the lattice matrix elements, using ePump~\cite{Schmidt:2018hvu,Hou:2019gfw} to update the PDFs.
We wish to test the sensitivity of the extracted gluon PDF to the data, so we perform a test, in which we increase the weight of the lattice data to see how the PDF changes.
We used the $a\approx 0.09$ fm RpITD from Ref.~\cite{Fan:2022kcb}, from which the lattice PDFs presented earlier in the study were extracted.
We show the results of this study in Fig.~\ref{fig:ePump_gluons}.
We can see that the direct RpITD information supports smaller gluon in the large $x$, and that increasing the weight continues to bring the mean value down further.
However, in any case, even with the strongest weight, the updated gluon PDFs are still consistent with the original CT18 PDF.
One source of this discrepancy, with the finding from the previous sections, could be the gluon momentum fraction.
Eq.~\eqref{eq:matching} suggests that the RpITD primarily gives information about the overall shape of the gluon PDF, but not the momentum fraction $\langle x_g\rangle$, while the PDFs studies in the earlier sections were the results of a full analysis combining RpITD and momentum fraction information.
Another feature to point out is that the relative errors on the PDFs are unchanged by the addition of the RpITD lattice data.

This is likely due to the relatively small number of lattice data points (35) compared to the number of experimental data points, $\mathcal{O}(1000)$, that are sensitive to the gluon PDF. 
This contrasts with the recent pion analysis of Ref.~\cite{Good:2025nny}, where 35 RpITD data points were included alongside only 108 experimental data points, most of which provided sensitivity to the gluon PDF only at small $x$.

Such a study also differs substantially from the previous JAM Collaboration study of polarized-gluon lattice matrix elements~\cite{Karpie:2023nyg}, for which the gluon helicity PDF extracted from existing experimental data has a significantly lower signal-to-noise ratio than its unpolarized counterpart. 
Nevertheless, the present study highlights an important challenge. 
While the experimental data provide strong overall constraints on the gluon PDF, they offer limited direct sensitivity in the region $x \in [0.3,0.8]$, where the lattice data are expected to be most informative. 
Treating the RpITD data on an equal footing with the much larger experimental data set may therefore dilute the impact of the lattice constraints, reducing their ability to improve the precision of the extracted gluon PDF in this region. 
Developing more effective strategies for combining lattice matrix elements with experimental measurements will require further investigation and remains an important direction for future work.

In summary, we find that the aforementioned approaches  yield different central values for the large-$x$ gluon PDF. 

Formally, both approaches are related through the same QCD factorization framework and should converge to the same gluon PDF  at sufficiently high statistical precision and with robust  treatment of the inverse problem. 
However, at the current level of statistical precision, the influence of the lattice RpITD data in the global fit appears to be overwhelmed by the much larger experimental data set.

Future analysis of the two methods at higher statistical precision, should be considered to determine more precisely the effects of the lattice data on the CTEQ-TEA  gluon PDF to eliminate any discrepancies between the two methodologies.
New algorithms and theoretical improvements will help us reach this higher statistical precision.
For example, recently proposed kinematically enhanced interpolating operators are expected to improve the precision of gluon RpITDs at high momentum, enabling more precise determinations of the gluon PDF within the pseudo-PDF framework~\cite{Zhang:2025hyo}. 
Furthermore, advanced machine-learning techniques have been proposed that could improve the precision of gluonic observables by factors of approximately $10$--$60$~\cite{Abbott:2026ylv}. The combination of these developments may enhance the statistical precision of lattice gluon data by as much as $\mathcal{O}(10$--$100)$. 
At such a level of precision, even  straightforward  inclusion of lattice nucleon gluon RpITD data in global fits alongside decades of experimental cross-section measurements  could yield tighter constraints on phenomenological determinations of the gluon PDF in the large-$x$ region.

\section{Conclusion}
\label{Sec:Conclusion}

In this work, we investigated the impact of the latest lattice gluon PDF~\cite{Fan:2022kcb,Good:2023gai} in the regions $x \in [0.3,0.8]$ and $x \in [0.4,0.7]$ on the CTEQ-TEA PDF series~\cite{Hou:2019efy,Hou:2022onq,Ablat:2024uvg}.
We explored the gluon PDF using two lattice spacings: the physical continuum limit ($a \to 0$) and $a \approx 0.09$~fm.
Our findings indicated that the $a \to 0$ dataset had a negligible impact on the PDFs, primarily due to the large systematic uncertainty in the continuum limit extrapolation.
Therefore, we focused on the $a \approx 0.09$~fm dataset, which represented the expected future uncertainty level in the QCD lattice calculations of the PDF nucleon gluon PDF.

In both the CT18~\cite{Hou:2019efy} and CT18As~\cite{Hou:2022onq} fits, the lattice dataset pushes the gluon central value to harder values for $x > 0.3$.
The corresponding uncertainty was significantly reduced, clearly demonstrating the constraining power of the lattice dataset in a global fit.
This complements experimental datasets in regions where precision was limited or absent.
As a trade-off, the intermediate $x$ gluon became slightly softer.
Due to the momentum sum rule, we observed a slight reduction in valence quark PDFs, such as the $d$ quark at large-$x$.
Comparing fits using CT18 and CT18As as bases, we found that the impact on the strange and anti-strange PDFs was parameterization dependent, affecting both the central value and the error band of the PDFs.
Furthermore, we observed that the lattice gluon PDF datasets in the regions $x \in [0.3, 0.8]$ and $x \in [0.4, 0.7]$ were largely consistent.
However, the more conservative Lat0.4to0.7 scenario showed a slightly smaller impact compared to the Lat0.3to0.8 scenario.

As the gluon PDF is largely constrained by collider jet datasets, we explored its interplay with the lattice dataset in Sec.~\ref{sec:interplay}.
Using the CT18 baseline, we directly compared the fit with the post-CT18 inclusive jet datasets (nIncJet)~\cite{Ablat:2024uvg} from ATLAS 8~TeV~\cite{ATLAS:2017kux}, 13~TeV~\cite{ATLAS:2017ble} and CMS 13~TeV~\cite{CMS:2021yzl}.
 Our analysis revealed that the lattice gluon dataset had a slightly greater impact on both the central value and the uncertainty band of the gluon PDF at large-$x$.
To further investigate the impact of jet and lattice datasets, we removed all collider jet datasets from the CT18 fit, obtaining a new fit referred to as CT18mJet.
We then compared this fit with those obtained by incorporating the lattice gluon dataset and the full set of collider jet datasets (AllJet) into the CT18mJet baseline.
In this scenario, we observed that AllJet had a significantly greater impact, largely attributable to its much larger datasets, comprising approximately 1000 data points.
Furthermore, while the lattice dataset primarily affected the gluon PDF in the large-$x$ region, the jet datasets exhibited a more extensive influence, constraining the gluon PDF across the entire $x$ range.
Our analysis indicates that all available jet data sets effectively constrain the gluon PDF uncertainty in the large $x$ region, and are consistent with results from lattice-QCD calculations.
The jet data currently utilized in the fit have been accumulated over decades, and collecting a comparable amount of new data would require a significant investment of time.
Furthermore, only jet measurements at high transverse momentum  $p_T$ and large rapidity region can effectively constrain the gluon PDF in the large $x$ region, where uncertainties remain substantial.
As a result, achieving precise control over PDF uncertainties in this region is particularly challenging.
In this study, however, we demonstrate that lattice QCD data can provide valuable constraints in the experimentally challenging large $x$ region, underscoring the importance of incorporating lattice input.
We also hope that lattice collaborations will extend their efforts to include distributions of
$u(\bar{u}), d(\bar{d}), s(\bar{s})$ or their combinations, which would greatly enhance our understanding of the proton’s internal structure and contribute to reducing PDF uncertainties.

To illustrate the phenomenological implications of the lattice gluon input, we examined the gluon-gluon luminosity $L_{gg}$, with the results presented in Fig.~\ref{Fig:PDFLuminosity}.
We observed an increase in the central value and a reduction in the uncertainty bands in the large invariant-mass region, naturally reflecting the corresponding impact on the large-$x$ gluon PDF.
As a phenomenological demonstration, we explored Higgs-like scalar production via gluon fusion at the LHC with a center-of-mass energy of 14~TeV.
By varying the scalar mass $M_H$, we observed a fully consistent behavior of the cross section with the $L_{gg}$ luminosity.
In a more realistic scenario, we considered the production of top-quark pairs, as well as their associated production with an additional Higgs boson, $Z$ boson, or jet, and analyzed the corresponding correlations.
We found a strong positive correlation among these processes, as they were all primarily driven by gluon-gluon fusion.
Furthermore, the inclusion of the lattice dataset increased their cross sections while simultaneously reducing the associated uncertainties.

Finally, we explored an alternative strategy for incorporating lattice-QCD information into global PDF analyses by directly fitting the gluon RpITD matrix elements within the pseudo-PDF framework, rather than first reconstructing an $x$-dependent PDF. 
While the PDF-based and RpITD-based approaches exhibit different preferences for the large-$x$ gluon central value, the resulting PDFs remain statistically consistent within current uncertainties. 
This observation is consistent with the expectation that both approaches should converge in the limits of high statistical precision and controlled systematic uncertainties. 
At present, however, the constraining power of the RpITD data appears to be limited by the relatively small number of lattice measurements compared to the much larger experimental dataset included in the global fit. 
These findings highlight both the promise of direct matrix-element constraints and the need for improved lattice precision and more sophisticated strategies for combining current lattice and experimental information in future analyses.

The significance of the lattice dataset extends beyond its direct impact on the gluon PDF, providing crucial constraints on the large-$x$ behavior of both gluons and quarks, as demonstrated in our previous study~\cite{Hou:2022onq} and this work.
By granting direct access to nonperturbative QCD effects, lattice calculations refine our understanding of PDFs in kinematic regions where traditional experimental constraints have been limited or ineffective.
This capability is particularly valuable for improving theoretical predictions in high-energy processes sensitive to large-$x$ dynamics, such as heavy resonance production and high momentum transfer scattering.
Incorporation of the lattice dataset into global PDF analyses enhances the precision of QCD studies and strengthens predictions for collider experiments, where large-$x$ uncertainties remain a significant challenge.
As lattice-QCD methodologies continue to advance, their integration into global fits will further deepen our understanding of partonic structure, enabling more precise and comprehensive descriptions of hadronic matter.

\begin{acknowledgements}

We would like to thank our CTEQ-TEA colleagues for many helpful discussions.
S.~Dulat was partly supported by the U.S. Department of Energy under Grant No.~DE-SC0010129.
S.~Dulat and A. Ablat were supported by the National Natural Science Foundation of China under Grant No.~11965020.
W.~Good was partially supported by the US National Science Foundation under grant PHY~2514533.
T.-J.~Hou was supported by the National Natural Science Foundation of China under Grant No.~12475079, and by the Natural Science Foundation of Hunan province of China under Grant No.~2023JJ30496.
K.~Xie and C.-P.~Yuan were supported by the U.S. National Science Foundation under Grants No.~PHY-2310291 and PHY-2310497.
H.~Lin was partially supported by the U.S. National Science Foundation under Grant No.~PHY~1653405 ``CAREER: Constraining Parton Distribution Functions for New-Physics Searches", Grant No.~PHY~2209424 and  PHY~2514533, and by the Research Corporation for Science Advancement through the Cottrell Scholar Award.

\end{acknowledgements}

\bibliographystyle{utphys}
\bibliography{ref}

\end{document}